\documentclass[twocolumn,numberedappendix]{openjournal}

\usepackage{caption}
\usepackage{float}
\usepackage{natbib}
\usepackage{xcolor}
\usepackage{textgreek}
\usepackage[utf8]{inputenc}
\usepackage[english]{babel}

\usepackage{hyperref}
\hypersetup{
    unicode, 
    colorlinks=true,
    linkcolor=linkcolor,
    citecolor=linkcolor,
    filecolor=linkcolor,
    urlcolor=linkcolor,
}
\usepackage{color}
\definecolor{linkcolor}{rgb}{0.0,0.3,0.5}
\usepackage{tensind}
\tensordelimiter{?}
\DeclareGraphicsExtensions{.bmp,.png,.jpg,.pdf}
\usepackage{verbatim}
\usepackage[normalem]{ulem}
\usepackage{orcidlink}
\usepackage{soul}

\graphicspath{ {./figs/} }

\begin{document}
\title{A Statistical--AI Framework for Detecting Transient Flares in SDSS Stripe 82 Quasar Light Curves}

\author{Atal Agrawal\altaffilmark{*}\orcidlink{0009-0009-8311-0523}}
\affiliation{Department of Physics, Indian Institute of Technology Roorkee, Roorkee 247667, India}
\altaffiltext{*}{E-mail: \href{mailto:atal_a@ph.iitr.ac.in}{atal\_a@ph.iitr.ac.in}}

\begin{abstract}
Quasars exhibit stochastic variability across wavelengths, typically well described by a Damped Random Walk (DRW). Occasionally, however, they undergo extreme luminosity changes---known as flares---that represent significant departures from this baseline behavior and provide valuable probes of accretion disc dynamics and the physics of supermassive black hole fueling. Although modern transient surveys have spurred growing interest in flare detection, no systematic search has yet been conducted within the legacy SDSS Stripe 82 dataset, which contains 9,258 spectroscopically confirmed quasars observed over a ${\sim}$10-year baseline. The principal statistical challenge is distinguishing these rare events from the ever-present stochastic variability. To address this, we present FLARE (Flare detection via physics-informed Learning, Anomaly scoring, and Recognition Engine), a modular three-stage framework for detecting transient flares in quasar light curves. FLARE models baseline DRW behavior, applies statistical anomaly scoring to flag candidate events, and employs a recognition engine to verify detections. For the Stripe 82 implementation, we deploy two complementary baselines---a physics-informed probabilistic Gated Recurrent Unit (GRU) trained on simulated DRW light curves, and an iterative Ornstein--Uhlenbeck (OU) process fitted directly to observed data with outlier masking---followed by Extreme Value Theory (EVT) for anomaly scoring. We benchmark twelve open-weight and proprietary Vision Language Models (VLMs) as recognition engines for final candidate verification. Detection is performed on $r$-band light curves, with candidates cross-checked against g-band data to rule out instrumental artifacts. Applying this framework, we identify 51 quasars exhibiting distinct flaring activity.
\end{abstract}

% Write your keywords here

\maketitle

\section{Introduction}
\label{sec:intro}

Quasars are extremely luminous active galactic nuclei (AGN) powered by supermassive black holes at their centers. They exhibit significant variability across the electromagnetic spectrum on timescales ranging from minutes to decades \citep{10.1093/mnras/stx1456}. This stochastic variability is well described by a one-dimensional Damped Random Walk (DRW) model \citep{2010ApJ...721.1014M}. Occasionally, however, quasars undergo extreme luminosity changes---known as flares---that represent significant departures from this baseline behavior. Such flares may be driven by enhanced black hole accretion, tidal disruption events \citep{Chanetal2019}, superluminous supernovae \citep{Drake:2011kg}, stellar-mass black hole mergers, or microlensing \citep{10.1093/mnras/stae1036}. Detecting and characterizing these events therefore provides valuable probes of the physical processes occurring in and around the accretion disc.
Several approaches have been employed to detect flares in quasar light curves, including sigma-clipping \citep{10.1093/mnras/stx1456}, Bayesian blocks combined with Gaussian Processes \citep{He_2025}, and Gaussian Processes alone \citep{10.1093/mnras/stae721}. However, the fundamental challenge remains: flares must be identified against an ever-present stochastically variable baseline, making it difficult to distinguish genuine transient events from rare but expected DRW fluctuations. At present, no single generalized framework exists that can be universally applied to detect flares across quasar datasets.

\begin{figure*}[!ht]
    \centering
    \includegraphics[width = \columnwidth]{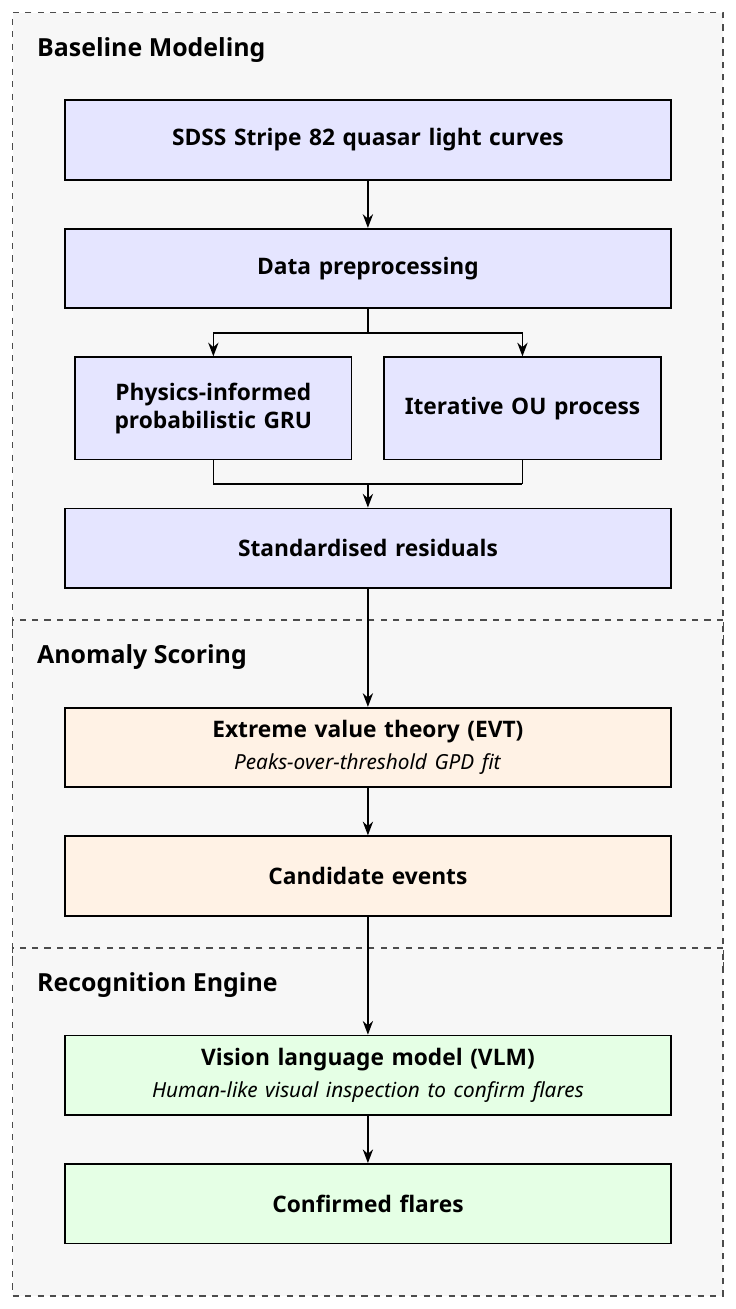}
    \caption{The FLARE framework. The pipeline consists of three stages: (1)~\textit{Baseline Modeling}, where a physics-informed probabilistic GRU or an iterative OU process models the DRW variability of each quasar light curve; (2)~\textit{Anomaly Scoring}, where standardized residuals are analyzed using Extreme Value Theory with a peaks-over-threshold GPD fit to flag candidate events; and (3)~\textit{Recognition Engine}, where Vision Language Models perform human-like visual inspection to confirm flare detections.}
    \label{fig:flareframework}
    \vspace{2mm}
\end{figure*}

To address this, we present FLARE (Flare detection via physics-informed Learning, Anomaly scoring, and Recognition Engine), a modular three-stage framework that models baseline DRW behavior, applies statistical anomaly scoring to flag candidate events, and employs a recognition engine to verify flare candidates. We apply FLARE to the SDSS Stripe 82 dataset, which covers $\sim 300 \text{ deg}^2$ on the celestial equator and contains $9258$ spectroscopically confirmed quasars observed over a ${\sim}\,10$-year baseline with ${\sim}\,60$--$80$ epochs per object---a temporal depth that newer surveys such as ZTF and the upcoming LSST are still building toward. Previously estimated DRW parameters \citep{2010ApJ...721.1014M} enable direct simulation of baseline behavior for each object. For the Stripe 82 implementation, we deploy two baselines---a physics-informed probabilistic Gated Recurrent Unit (GRU) and an iterative Ornstein--Uhlenbeck (OU) process---followed by Extreme Value Theory (EVT) for anomaly scoring and Vision Language Models (VLMs) as recognition engines. Applying this framework, we identify 51 quasars exhibiting distinct flaring activity.
The remainder of this paper is organized as follows. In Section~\ref{sec:framework}, we describe the FLARE framework. Section~\ref{sec:data} presents the data and its preprocessing. Section~\ref{sec:methods} and Section~\ref{sec:results} detail the method and results, followed by discussion and conclusions in Section~\ref{sec:discussions} and Section~\ref{sec:conclusion}. The light curves of all 51 confirmed flaring quasars are presented in the Appendix~\ref{app:flares}.

\section{FLARE FRAMEWORK}
\label{sec:framework}

% \begin{figure}[H]
%     \centering
%     \setlength{\fboxsep}{0pt}
%     \setlength{\fboxrule}{0.5pt}
%     \fbox{\includegraphics[width=0.8\columnwidth]{figs/REngine.pdf}}
%     \caption{Architecture of the recognition engine. Two VLMs act as independent classifiers, each providing a flare/non-flare classification for the candidate light curve. A third VLM serves as an evaluator, flagging misclassifications and providing feedback to the classifiers. This evaluation--feedback cycle is repeated for two iterations to allow the classifiers to refine their predictions.}
%     \label{fig:vlm_engine}
% \end{figure}

\begin{figure}[!ht]
    \centering
    \includegraphics[width=\columnwidth]{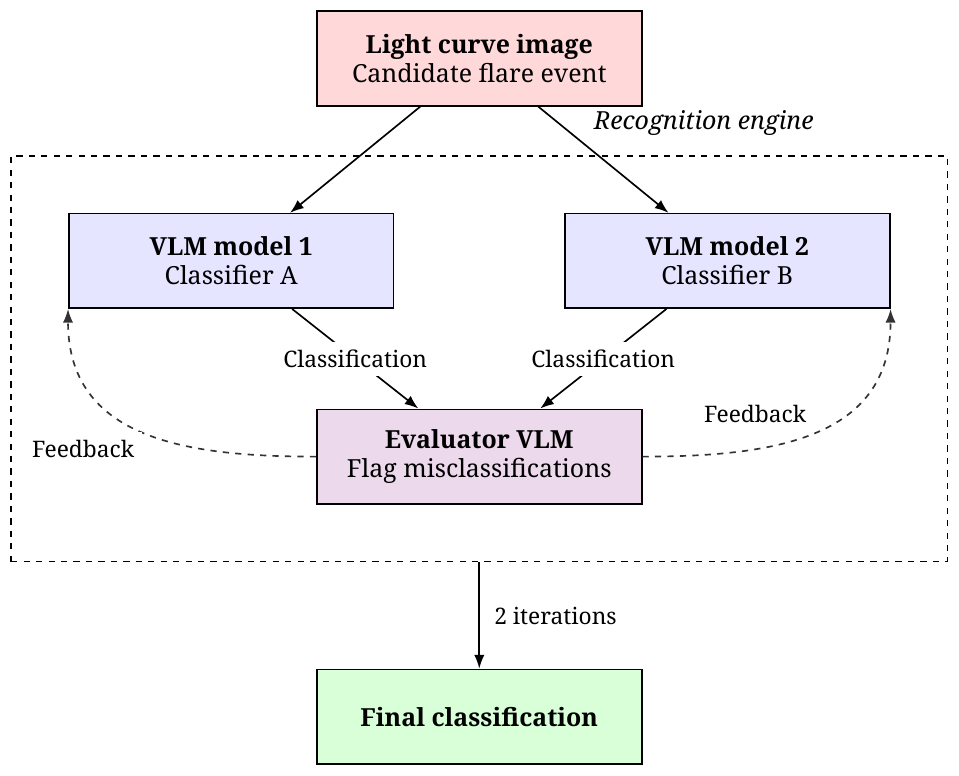}
    \caption{Architecture of the recognition engine. Two VLMs act as independent classifiers, each providing a flare/non-flare classification for the candidate light curve. A third VLM serves as an evaluator, flagging misclassifications and providing feedback to the classifiers. This evaluation--feedback cycle is repeated for two iterations to allow the classifiers to refine their predictions.}
    \label{fig:vlm_engine}
    \vspace{2mm}
\end{figure}

Detecting flares in quasar light curves is fundamentally an anomaly detection problem. The key steps are: defining the baseline behavior, flagging anomalous deviations, and inspecting the flagged candidates to confirm they are genuine anomalies. Along these lines, we present the FLARE framework (Figure~\ref{fig:flareframework}), which operates in three stages: baseline modeling, anomaly scoring, and candidate verification through a recognition engine.

The framework is designed to be modular. For the baseline modeling stage, any suitable model for DRW variability can be employed, such as Gaussian Processes \citep{10.1093/mnras/stae721} or comparison of DRW parameters derived from different baseline lengths to identify objects whose variability properties have changed significantly \citep{Suberlak2021}. For anomaly scoring, methods such as sigma-clipping on de-trended light curves \citep{10.1093/mnras/stx1456} or Bayesian block segmentation combined with Gaussian Processes \citep{He_2025} can be used to flag statistically significant deviations. The recognition engine then verifies candidates through automated visual classification.

Figure~\ref{fig:vlm_engine} shows the architecture of the recognition engine. It employs three VLMs: two as primary classifiers (A and B) and a third as an evaluator. The two classifiers are selected to complement each other---one optimized for high recall to maximize detection of genuine flares, and the other for high precision to minimize false positives. The evaluator requires high overall accuracy, as it must correctly adjudicate both missed flares and false detections from the classifiers. The pipeline operates over two rounds. In Round~1, both classifiers independently examine each candidate light curve image and provide a flare/non-flare classification. When both classifiers agree, the classification is accepted. When they disagree, the evaluator reviews the light curve image alongside both classifier outputs and makes a final determination. The evaluator then generates actionable feedback summarizing the error patterns observed in the disputed cases. In Round~2, the disputed cases from Round~1 are reclassified by both classifiers, now augmented with the evaluator's feedback. Cases where the classifiers now agree are accepted; any remaining disagreements are resolved by a second evaluator pass, whose verdict is final. We benchmark 12 VLMs for the classifier and evaluator roles, which we discuss in Section~\ref{sec:methods}.

\section{Data and Simulations}
\label{sec:data}

We work with the SDSS Stripe 82 quasar light curve data compiled by \citet{2010ApJ...721.1014M}, who fitted an OU process to 9,258 spectroscopically confirmed quasars, providing DRW parameters ($\tau$, $\hat{\sigma}$) for each object.
We use the $r$-band photometry, as it offers the best combination of photometric depth and cadence in the Stripe 82 dataset, and lies near the peak of the quasar spectral energy distribution at the typical redshifts of this sample, maximizing sensitivity to intrinsic luminosity variations. The data are preprocessed by first removing bad observations flagged as $-99.99$ or $99.99$ in the catalog and correcting for Galactic extinction using the values provided in the S82 QSO data file. We then remove single-point spikes using a Median Absolute Deviation (MAD) based continuity check: for each interior point $i$, we compute the expected magnitude as the mean of its neighbors,
\begin{equation}
\hat{m}_i = \frac{m_{i-1} + m_{i+1}}{2},
\end{equation}
and remove the point if
\begin{equation}
|m_i - \hat{m}_i| > 5\,\sigma_{\mathrm{MAD}},
\end{equation}
where $\sigma_{\mathrm{MAD}} = \mathrm{MAD}/0.6745$ is the robust standard deviation estimated from the median absolute deviation of the light curve magnitudes. The first and last points of each light curve are retained.

\subsection{Simulated DRW Light Curves}

For baseline modeling with the GRU and for VLM benchmarking, we simulate DRW light curves by drawing realizations from a Gaussian process with the DRW kernel \citep{2009ApJ...698..895K},
\begin{equation}
k(\Delta t) = \hat{\sigma}^2 \, \exp\!\left(-\frac{\Delta t}{\tau}\right),
\end{equation}
where $\tau$ is the damping timescale and $\hat{\sigma}$ is the variability amplitude, both taken from the fitted parameters of \citet{2010ApJ...721.1014M}. The realization is evaluated at the same MJD timestamps as the observed light curve, preserving the cadence and sparsity of the Stripe 82 survey. The simulated magnitude is then shifted to match the mean observed magnitude $\bar{m}$ of each object, and Gaussian noise is injected using the per-epoch photometric errors $\epsilon_i$ from the original light curve:
\begin{equation}
m_{\mathrm{sim},i} = m_{\mathrm{DRW},i} + \bar{m} + \mathcal{N}(0,\, \epsilon_i),
\end{equation}

where $m_{\mathrm{DRW},i}$ is the zero-mean DRW realization at epoch $i$. We generate seven independent sets of 9,258 simulated light curves using different random seeds: one for training the baseline model, one for validation, and five for benchmarking the recognition engine VLMs.

\subsection{Synthetic Flare Injection}

To benchmark the recognition engine, we inject synthetic flares into the simulated DRW light curves. All flare injections are performed in flux space to ensure physically consistent magnitude changes. The baseline magnitude is first converted to flux via
\begin{equation}
F_{\mathrm{base}} = 10^{-0.4\,m},
\end{equation}
and the peak flare flux is computed from a desired brightening amplitude $A$ (in magnitudes) as
\begin{equation}
F_{\mathrm{peak}} = \tilde{F}_{\mathrm{base}} \left(10^{0.4\,A} - 1\right),
\end{equation}
where $\tilde{F}_{\mathrm{base}}$ is the median baseline flux. A temporal profile $\phi(t)$, normalized to unit peak, is scaled by $F_{\mathrm{peak}}$ and added to the baseline flux. The total flux is then converted back to magnitude:
\begin{equation}
m_{\mathrm{flare}}(t) = -2.5\,\log_{10}\!\left(F_{\mathrm{base}}(t) + F_{\mathrm{peak}}\;\phi(t)\right).
\end{equation}

\begin{figure}[H]
    \centering
    \setlength{\fboxsep}{0pt}
    \setlength{\fboxrule}{0.5pt}
    \fbox{\includegraphics[width=0.7\columnwidth]{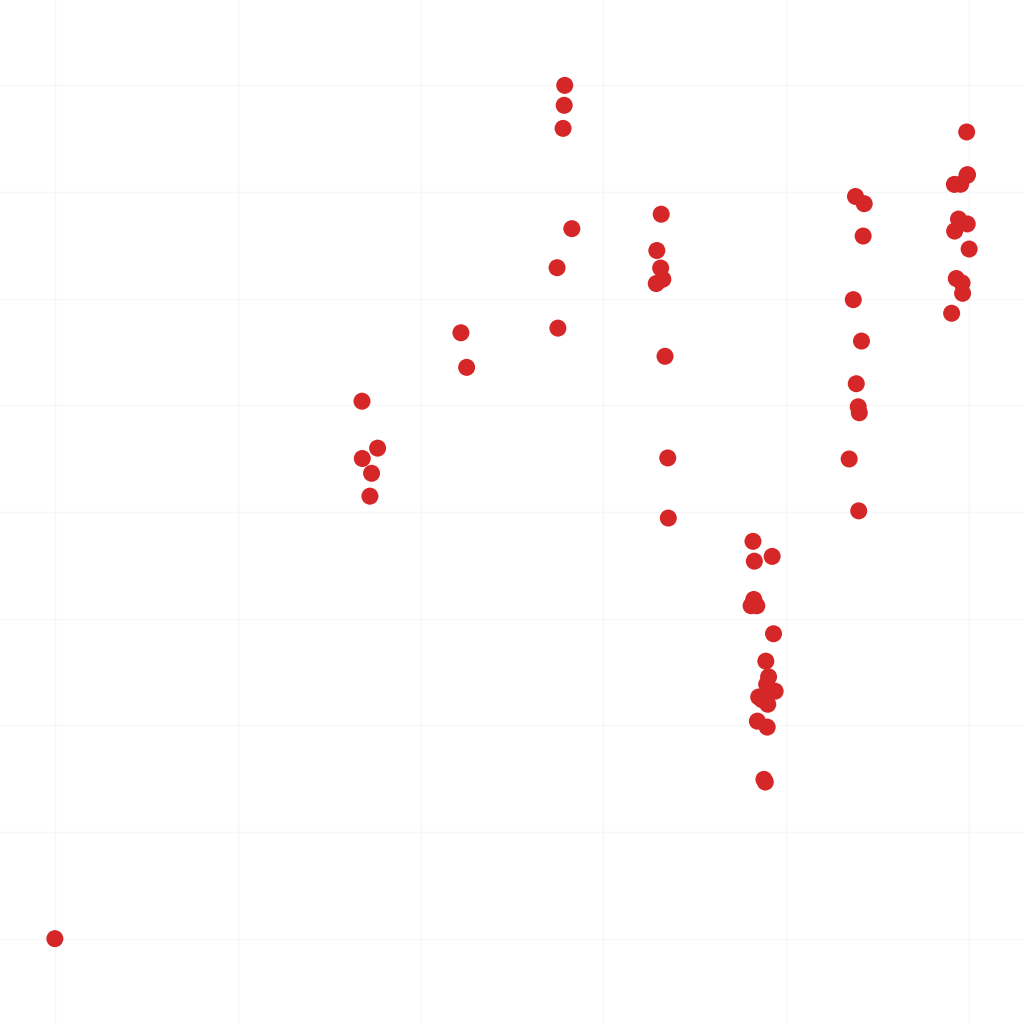}}
    \caption{Example simulated DRW light curve with Gaussian noise injected based on per-epoch Stripe~82 photometric errors. Axes and labels are intentionally omitted as these images represent the direct morphological inputs fed to the VLMs.}
    \label{fig:drw}
\end{figure}

We inject three morphologically distinct flare types. For all three, the peak time $t_{\mathrm{peak}}$ is drawn randomly from the observed MJD timestamps, ensuring that the flare peak coincides with an observed epoch. The amplitude is drawn uniformly from $A \in [0.3,\, 1.2]$~mag.

\subsubsection{FRED (Fast Rise Exponential Decay)}

The FRED profile is defined as
\begin{equation}
\phi(t) =
\left\{
\begin{array}{ll}
e^{(t - t_{\mathrm{peak}})/\tau_{\mathrm{rise}}}, & t < t_{\mathrm{peak}}, \\[4pt]
e^{-(t - t_{\mathrm{peak}})/\tau_{\mathrm{decay}}}, & t \geq t_{\mathrm{peak}},
\end{array}
\right.
\end{equation}
with $\tau_{\mathrm{rise}} \in [10,\, 50]$~days and $\tau_{\mathrm{decay}} \in [100,\, 400]$~days.

\subsubsection{Gaussian}

The Gaussian profile is
\begin{equation}
\phi(t) = \exp\!\left(-\frac{(t - t_{\mathrm{peak}})^2}{2\,\sigma_{\mathrm{flare}}^2}\right),
\end{equation}
with $\sigma_{\mathrm{flare}} \in [20,\, 120]$~days.

\subsubsection{Gamma}

The Gamma profile is
\begin{equation}
\phi(t) \propto (t - t_0)^{k-1}\;\exp\!\left(-\frac{t - t_0}{\theta}\right), \quad t > t_0,
\end{equation}
normalized to unit peak, where $t_0 = t_{\mathrm{peak}} - (k-1)\,\theta$ ensures the peak occurs at $t_{\mathrm{peak}}$, with shape parameter $k \in [2,\, 5]$ and timescale $\theta \in [20,\, 100]$~days.

\begin{figure}[H]
    \centering
    \setlength{\fboxsep}{0pt}
    \setlength{\fboxrule}{0.5pt}
    \fbox{\includegraphics[width=0.7\columnwidth]{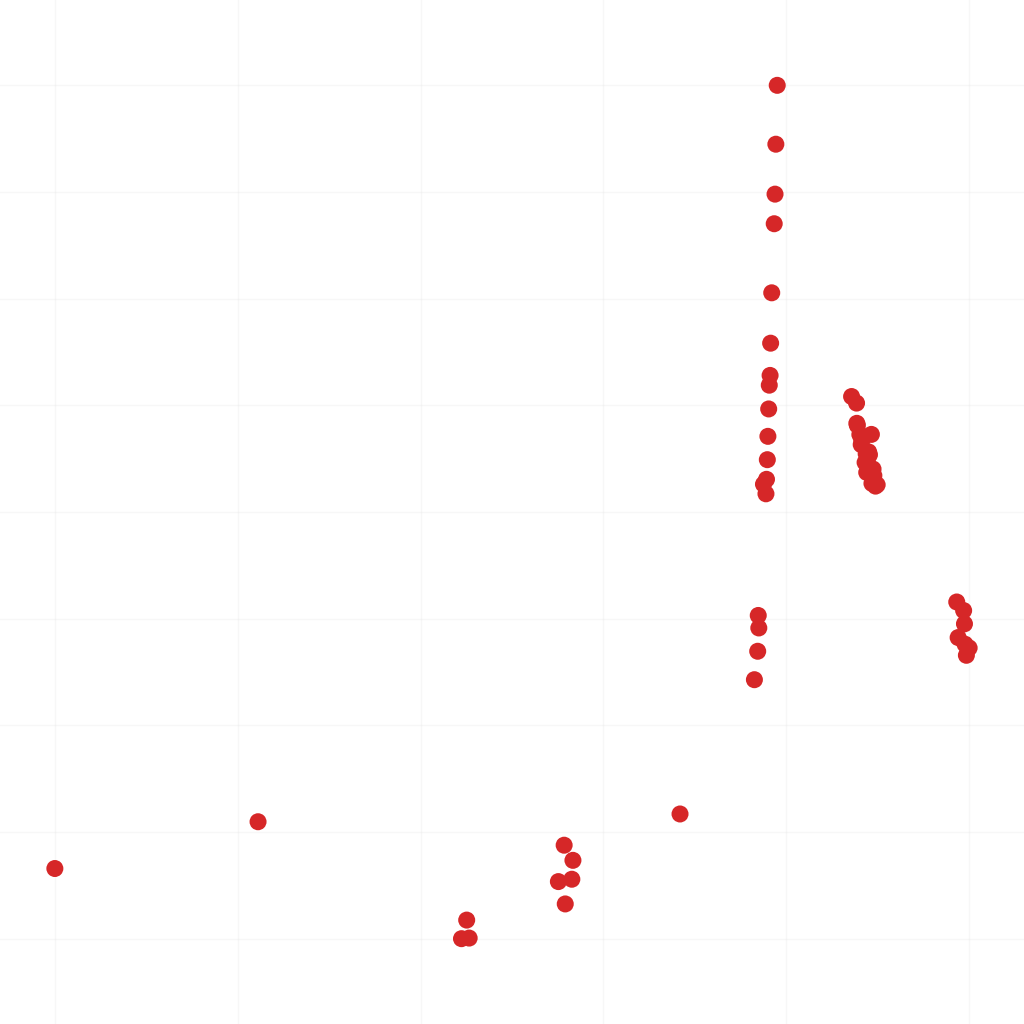}}
    \caption{Example simulated DRW light curve with an injected FRED flare. Axes and labels are intentionally omitted as these images represent the direct morphological inputs fed to the VLMs.}
    \label{fig:Fred}
\end{figure}

\subsection{Single-Point Spike Injection}

In addition to the three flare classes, we inject single-point spikes into a fourth set of simulated light curves to test whether the VLMs can distinguish genuine multi-epoch flares from isolated photometric artifacts. Single-point spikes are treated as non-flare events, as they represent isolated photometric artifacts rather than astrophysical transients. A single epoch $j$ is selected randomly from the observed timestamps and brightened in flux space:
\begin{equation}
F_{\mathrm{spike}} = \tilde{F}_{\mathrm{base}} \left(10^{0.4\,A} - 1\right),
\end{equation}
added only at epoch $j$:
\begin{equation}
m_{\mathrm{spike},i} =
\left\{
\begin{array}{ll}
-2.5\,\log_{10}\!\left(F_{\mathrm{base},i} + F_{\mathrm{spike}}\right), & i = j, \\[4pt]
m_i, & i \neq j,
\end{array}
\right.
\end{equation}
with the amplitude drawn uniformly from $A \in [0.3,\, 1.5]$~mag.
\vspace{-2mm}

\begin{figure}[H]
    \centering
    \setlength{\fboxsep}{0pt}
    \setlength{\fboxrule}{0.5pt}
    \fbox{\includegraphics[width=0.7\columnwidth]{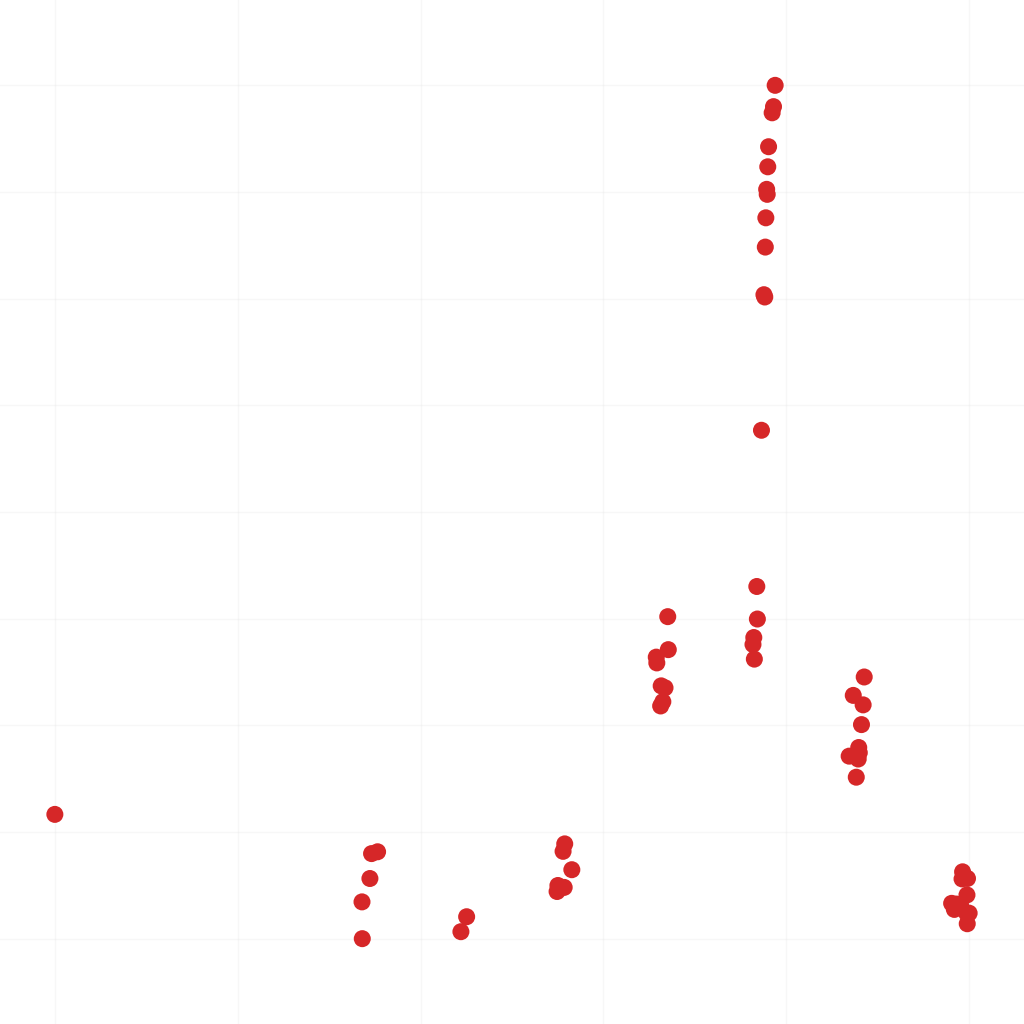}}
    \caption{Example simulated DRW light curve with an injected Gamma flare. Axes and labels are intentionally omitted as these images represent the direct morphological inputs fed to the VLMs.}
    \label{fig:Gamma}
\end{figure}

\subsection{Benchmarking Dataset}
The benchmarking dataset comprises five classes: three flare types (FRED, Gaussian, Gamma), pure DRW (no injection), and single-point spikes. We simulate four sets of DRW light curves for benchmarking, injecting one flare type into each: FRED, Gaussian, Gamma, and single-point spikes respectively. A fifth set of pure DRW light curves serves as the baseline representing normal quasar variability. This five-class design mitigates the ${\sim}\,50\%$ baseline accuracy that a binary flare/non-flare classification would yield under random guessing, requiring the VLMs to distinguish between morphologically distinct event types.

\begin{figure}[H]
    \centering
    \setlength{\fboxsep}{0pt}
    \setlength{\fboxrule}{0.5pt}
    \fbox{\includegraphics[width=0.7\columnwidth]{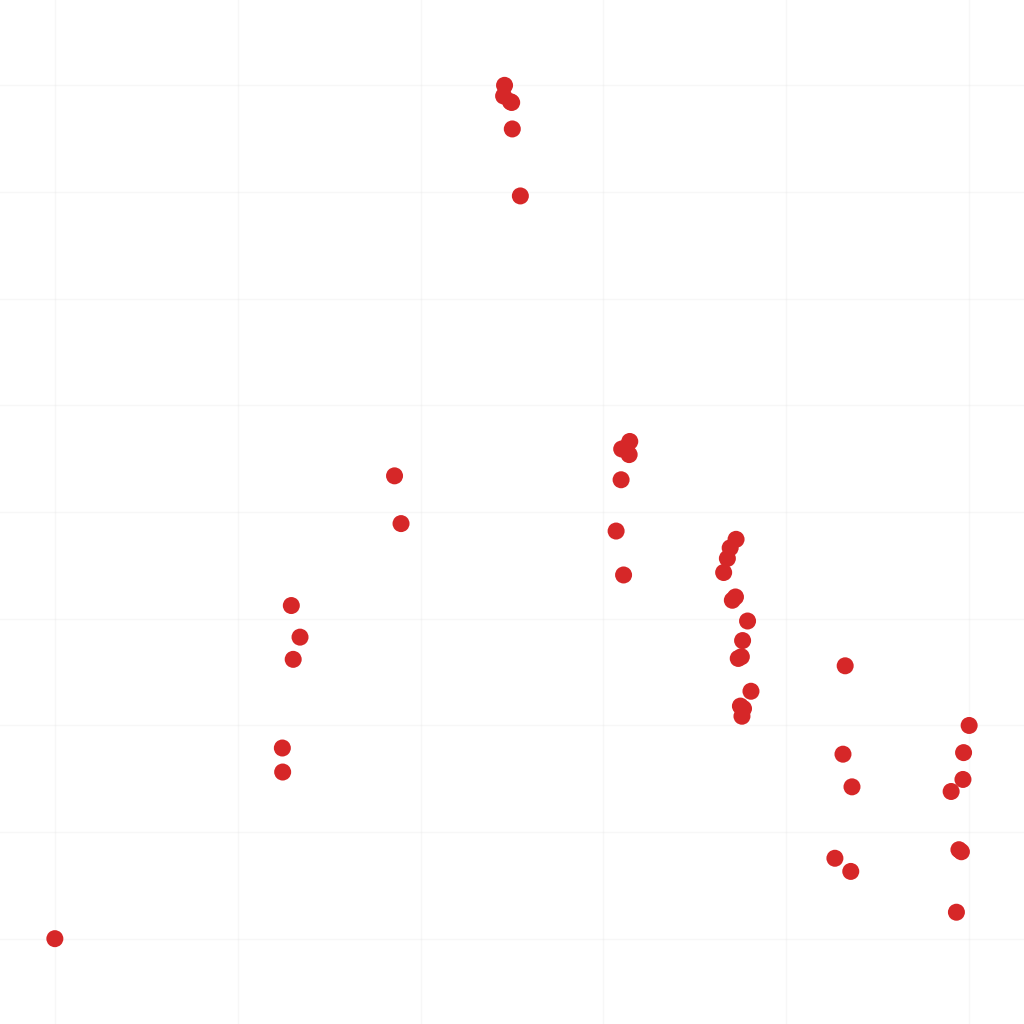}}
    \caption{Example simulated DRW light curve with an injected Gaussian flare. Axes and labels are intentionally omitted as these images represent the direct morphological inputs fed to the VLMs.}
    \label{fig:Gaussian}
\end{figure}

\begin{figure}[H]
    \centering
    \setlength{\fboxsep}{0pt}
    \setlength{\fboxrule}{0.5pt}
    \fbox{\includegraphics[width=0.7\columnwidth]{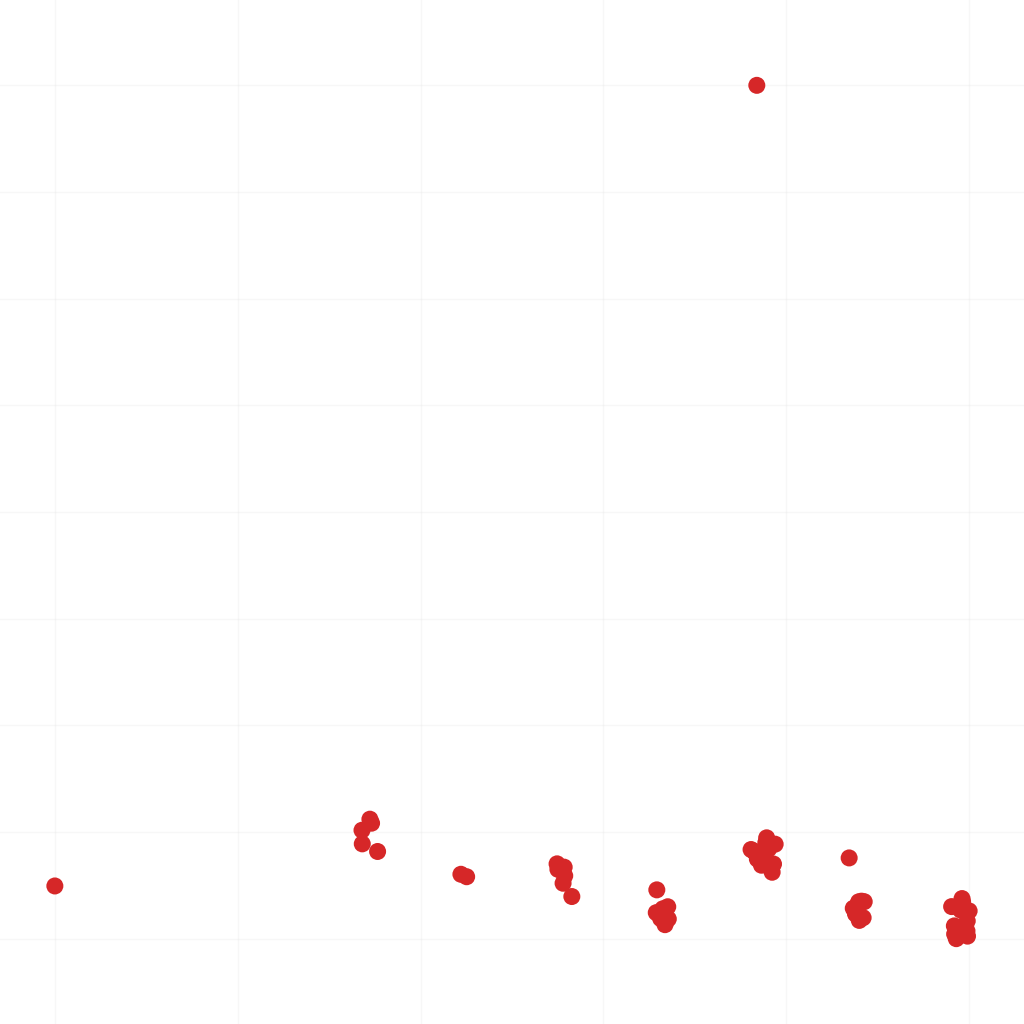}}
    \caption{Example simulated DRW light curve with an injected single-point spike. Axes and labels are intentionally omitted as these images represent the direct morphological inputs fed to the VLMs.}
    \label{fig:spike}
\end{figure}

Figures~\ref{fig:drw}, \ref{fig:Fred}, \ref{fig:Gamma}, \ref{fig:Gaussian}, and \ref{fig:spike} show examples of simulated light curves for each of the five classes. The plots use a white background with a faint grid and no axis labels, so that the VLMs classify based solely on the morphology of the light curve without being biased by time or magnitude scales. Since each set is simulated on the timestamps of the Stripe~82 data, each contains ${\sim}\,9{,}260$ light curves, giving a total of ${\sim}\,46{,}300$ across all five classes. We split this into 80\% training, 10\% validation, and 10\% test data, ensuring that each of the five classes is proportionally represented in all three splits. The training and validation splits are used for parameter-efficient fine-tuning of an open-weight VLM, discussed in Section~\ref{sec:methods}. For benchmarking, we use the same test set for all 12 VLMs, comprising 4,630 light curves with 926 per class.

\section{Methods}
\label{sec:methods}

\subsection{Baseline Modeling}

Modeling the stochastic variability of quasars is essential for identifying anomalous departures such as flares. The simplest widely used model is the DRW \citep{2009ApJ...698..895K,2010ApJ...721.1014M}, equivalent to the lowest-order continuous-time autoregressive moving-average (CARMA) process, CARMA(1,0). Higher-order CARMA models \citep{Kelly2014} offer greater flexibility and can capture more complex power spectral density shapes, with studies finding that many quasar light curves prefer orders around $(p,q) \approx (4,2)$ \citep{Yu2022}. More recently, latent stochastic differential equations have been applied to jointly model quasar variability and accretion disk reprocessing \citep{Fagin2024}, and recurrent autoencoder architectures have been used for variability-based classification \citep{Tachibana2020}.

While these approaches offer richer representations of quasar variability, they introduce additional free parameters that risk overfitting the very transients we seek to detect---a flexible model may absorb a flare into its baseline fit rather than leaving it as a residual. For flare detection, a constrained OU baseline is therefore preferable: because it cannot accommodate transient events within its fit, they are preserved as large residuals in the output. This motivates our choice of OU-based baselines for both arms of the pipeline.

We employ two baselines. The first is a physics-informed probabilistic Gated Recurrent Unit (GRU) trained on simulated DRW light curves, which learns a data-driven representation of baseline variability regularized by the known OU process dynamics. The second is an iterative OU process fitted directly to the observed data using the \texttt{celerite} Gaussian process library \citep{celerite}, with outlier masking to obtain a robust baseline even in the presence of flares. These two approaches have complementary sensitivity profiles: the GRU, trained on simulated data, enforces a rigid baseline and is sensitive to transients that a flexible model might absorb into its fit, while the iterative OU process adapts to the real data with a lower detection threshold and captures subtler deviations. Neither baseline alone is sufficient; their union provides a more complete flare census.

\subsubsection{Physics-Informed Probabilistic GRU}

\begin{figure*}[!ht]
    \centering
    \includegraphics[width=\textwidth]{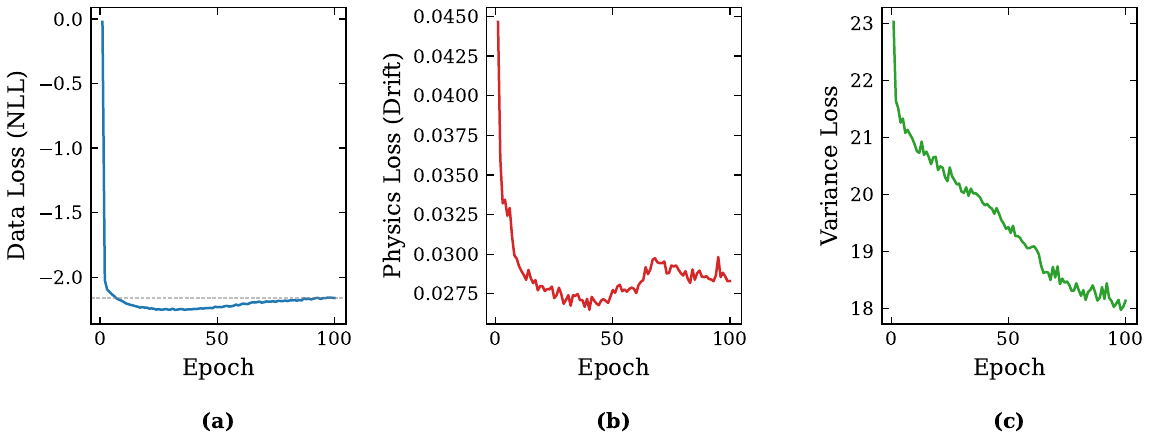}
    \caption{Training loss curves for the physics-informed probabilistic GRU over 100 epochs. (a)~Data loss (negative log-likelihood), which converges within the first ${\sim}\,20$ epochs. (b)~Physics loss (drift regularizer), measuring deviation from the OU conditional mean. (c)~Variance loss, measuring deviation from the OU conditional variance. All three components are shown unweighted; the annealed weights $\lambda_{\mathrm{phys}}$ and $\lambda_{\mathrm{var}}$ increase linearly over training, gradually enforcing physical consistency.}
    \label{fig:losscurve}
    \vspace{2mm}
\end{figure*}

We choose a GRU architecture over alternative sequence models for several reasons. Standard RNNs lack gating mechanisms, making them less effective at selectively retaining information across irregularly spaced time steps. Long Short-Term Memory (LSTM) networks address this but introduce three gates and a separate cell state, roughly doubling the parameter count relative to a GRU for comparable performance. Transformers, while powerful, require quadratic memory in sequence length, lack an inherent notion of irregular time spacing without additional positional encoding schemes, and their global self-attention mechanism captures long-range dependencies between distant epochs, whereas DRW baseline prediction is fundamentally a short-range task---each predicted magnitude depends primarily on the immediately preceding observation and the time elapsed since then. The GRU offers an effective compromise: its reset and update gates selectively control information flow, it has fewer parameters than an LSTM, naturally accommodates irregular cadence when $\Delta t_i$ is provided as an explicit input feature, and its recurrent structure inherently prioritizes recent context, aligning well with the Markovian nature of the OU process.

For baseline modeling of DRW variability in the Stripe~82 data, the GRU takes as input the mean-centered magnitude $m_i$ and the time step $\Delta t_i = t_i - t_{i-1}$ at each epoch, and outputs a predictive mean $\mu_i$ and uncertainty $\sigma_i$ for the next observation. The predicted uncertainty is parameterized via a softplus activation with a floor of $0.02$~mag and clamped to $[0.02,\, 5.0]$~mag to ensure numerical stability. Input magnitudes are mean-centered per object prior to training.

The model is trained on simulated DRW light curves (Section~\ref{sec:data}) using a composite loss function inspired by the physics-informed neural network (PINN) framework \citep{Raissi2019}, comprising three terms:
\begin{equation}
\mathcal{L} = \mathcal{L}_{\mathrm{data}} + \lambda_{\mathrm{phys}}\,\mathcal{L}_{\mathrm{drift}} + \lambda_{\mathrm{var}}\,\mathcal{L}_{\mathrm{var}}.
\end{equation}
The first term is the negative log-likelihood (NLL) of a Gaussian predictive distribution:
\begin{equation}
\mathcal{L}_{\mathrm{data}} = \frac{1}{N}\sum_i \left[\frac{1}{2}\log\sigma_i^2 + \frac{(m_i - \mu_i)^2}{2\,\sigma_i^2}\right].
\end{equation}
The second term is a physics-informed drift regularizer that penalizes deviations from the expected OU conditional mean:
\begin{equation}
\mathcal{L}_{\mathrm{drift}} = \frac{1}{N}\sum_i \frac{(\mu_i - \mu_{\mathrm{OU},i})^2}{\sigma^2_{\mathrm{OU},i}},
\end{equation}
where the OU conditional mean and variance \citep{2009ApJ...698..895K} are
\begin{equation}
\mu_{\mathrm{OU},i} = \bar{m} + (m_{i-1} - \bar{m})\,e^{-\Delta t_i / \tau},
\end{equation}
\begin{equation}
\sigma^2_{\mathrm{OU},i} = \frac{\hat{\sigma}^2 \tau}{2}\left(1 - e^{-2\Delta t_i / \tau}\right).
\end{equation}

The third term is a variance regularizer that encourages the model's predicted variance to match the OU expected variance:
\begin{equation}
\mathcal{L}_{\mathrm{var}} = \frac{1}{N}\sum_i \left(\sigma_i^2 - \sigma^2_{\mathrm{OU},i}\right)^2.
\end{equation}

\begin{figure*}[!ht]
    \centering
    \includegraphics[width=\textwidth]{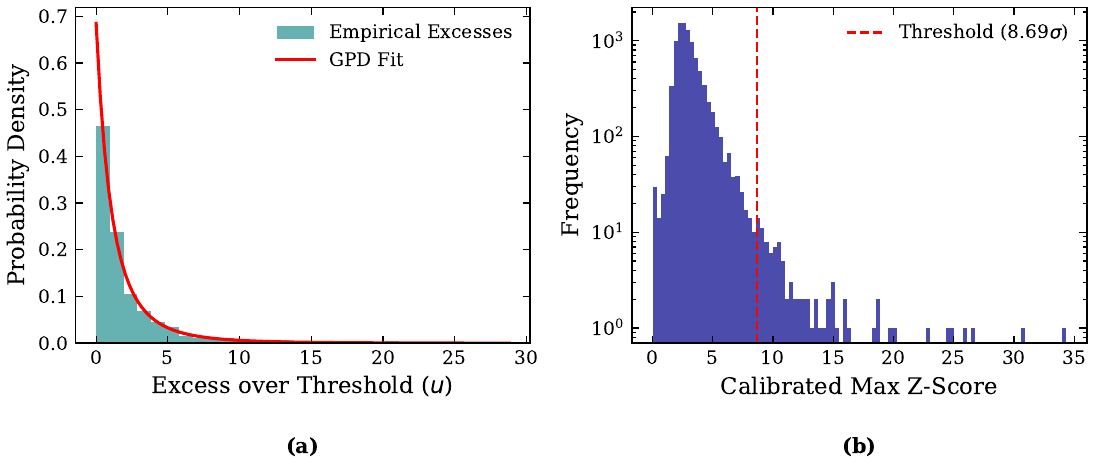}
    \caption{EVT anomaly scoring for the GRU baseline. (a)~GPD fit to the empirical exceedances above the 95th percentile threshold. (b)~Distribution of calibrated maximum $|z|$-scores across 9,258 simulated quasars, with the detection boundary at $8.69\sigma$ (dashed line). Objects exceeding this threshold yield 51 candidates.}
    \label{fig:evt_gru}
    \vspace{2mm}
\end{figure*}
The regularization weights $\lambda_{\mathrm{phys}}$ and $\lambda_{\mathrm{var}}$ are linearly annealed over training \citep{Bengio2009}, from 0.005 to 0.05 and from 0.01 to 0.20 respectively, allowing the model to first learn the data distribution before gradually enforcing consistency with the OU process. The GRU has a hidden dimension of 64 and a linear output head mapping to two parameters ($\mu_i$, $\sigma_i$). Each light curve is processed individually with a batch size of 1. The model is trained for 100 epochs using the Adam optimizer with a learning rate of $10^{-3}$ and gradient clipping at 1.0. Training loss curves are shown in Figure~\ref{fig:losscurve}. The data loss converges within the first ${\sim}\,20$ epochs, and the drift loss decreases steadily, indicating that the GRU's predictions align with the OU conditional mean. The variance loss decreases but does not fully converge, as the variance regularization acts as a soft constraint rather than enforcing exact agreement with the OU conditional variance. The data loss and drift loss are the primary convergence indicators.

\subsubsection{Iterative OU Process}
\label{sec:iter_ou}
The second baseline fits an OU process directly to the observed $r$-band light curve using the \texttt{celerite} Gaussian process library \citep{celerite}, iteratively masking outliers to obtain robust DRW parameters even in the presence of flares. Direct OU fitting to observed data has the advantage of not relying on externally derived DRW parameters, which may be biased if the light curve baseline is shorter than ${\sim}\,10\tau$ \citep{Kozlowski2017,Suberlak2021}. By fitting directly and iteratively masking outliers, the procedure yields DRW parameters that are self-consistent with the unmasked data, ensuring that the baseline is not contaminated by the very transients we seek to detect.
The kernel is a sum of a DRW term and a jitter term:
\begin{equation}
k(\Delta t) = a\,\exp(-c\,\Delta t) + \sigma_{\mathrm{jit}}^2\,\delta(\Delta t),
\end{equation}
where $a$ and $c$ are related to the DRW parameters via $\hat{\sigma} = \sqrt{a}$ and $\tau = 1/c$, and $\sigma_{\mathrm{jit}}$ accounts for excess white noise beyond the reported photometric errors. The hyperparameters are optimized by maximizing the GP log-likelihood using L-BFGS-B, with the mean magnitude fitted simultaneously.

Given the fitted parameters $(\tau, \hat{\sigma}, \bar{m})$, we compute one-step-ahead OU conditional predictions for each epoch. The conditional mean and variance at epoch $i$, given the observed magnitude $m_{i-1}$, are
\begin{equation}
\mu_{\mathrm{OU},i} = \bar{m} + (m_{i-1} - \bar{m})\,e^{-\Delta t_i / \tau},
\end{equation}
\begin{equation}
\sigma^2_{\mathrm{OU},i} = \frac{\hat{\sigma}^2\,\tau}{2}\left(1 - e^{-2\Delta t_i / \tau}\right),
\end{equation}
where $\Delta t_i = t_i - t_{i-1}$. The standardized residual at each epoch is then
\begin{equation}
z_i = \frac{m_i - \mu_{\mathrm{OU},i}}{\sqrt{\sigma^2_{\mathrm{OU},i} + \epsilon_i^2}},
\label{eq:ou_resid}
\end{equation}
where $\epsilon_i$ is the photometric error at epoch $i$. The inclusion of $\epsilon_i^2$ in the denominator ensures that the residuals are calibrated against the total predictive uncertainty---both intrinsic OU variance and measurement noise.

The iterative masking procedure operates as follows. At each iteration, all epochs with $|z_i| > z_{\mathrm{mask}}$ are flagged, excluding the first and last epochs which are always retained. The DRW parameters are then re-estimated on the unmasked data, and new residuals are computed for the full light curve. This cycle repeats until the mask converges (i.e., no new points are flagged or unflagged) or a maximum of 10 iterations is reached. Light curves with fewer than 15 unmasked epochs are excluded. The procedure converges for 9,104 of the 9,258 quasars. We adopt $z_{\mathrm{mask}} = 3.0$ as the default. The per-object summary statistic passed to the anomaly scoring stage is $\max_i |z_i|$, the maximum absolute standardized residual across all epochs.
\begin{figure*}[!ht]
    \centering
    \includegraphics[width=\textwidth]{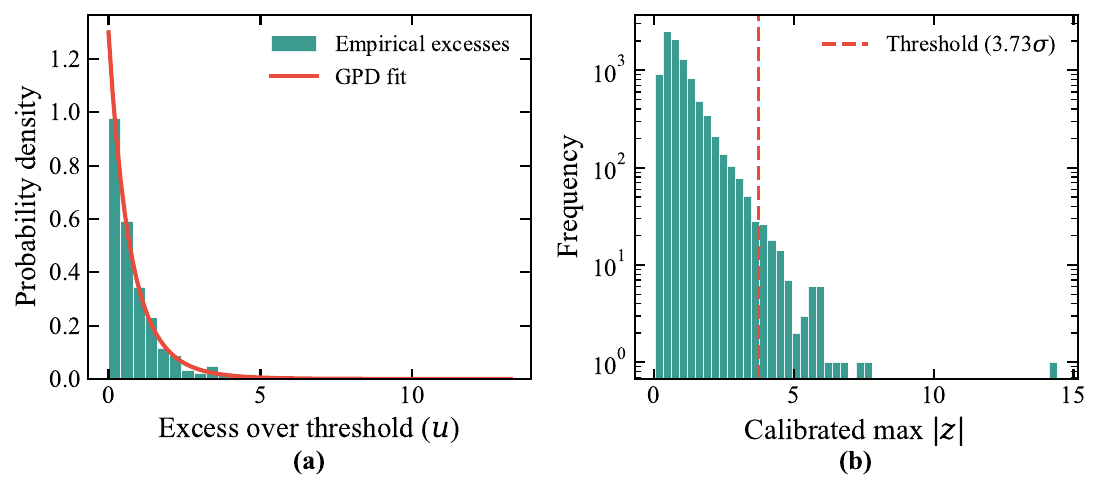}
    \caption{EVT anomaly scoring for the iterative OU baseline. (a)~GPD fit to the empirical exceedances above the 95th percentile threshold. (b)~Distribution of maximum $|z|$-scores across 9,104 converged quasars, with the detection boundary at $3.73\sigma$ (dashed line), yielding 92 candidates.}
    \label{fig:evt_ou}
    \vspace{2mm}
\end{figure*}

\subsection{Anomaly Scoring: Extreme Value Theory}

We use Extreme Value Theory (EVT) to derive principled detection thresholds for both baselines, rather than relying on fixed sigma-clipping. Within EVT, two approaches are commonly used: Block Maxima and Peaks-Over-Threshold (POT). Block Maxima requires partitioning each light curve into fixed-length blocks and extracting the maximum from each. However, the Stripe~82 data are sparsely sampled (${\sim}\,60$--$80$ epochs over ${\sim}\,10$ years), and the timescale of a potential flare is unknown a priori---a flare could span a duration longer than any reasonable block size, leading to loss of information. The POT approach avoids this limitation by modeling all exceedances above a threshold $u$, making it better suited for sparse, irregularly sampled data. We therefore adopt the POT method for both baselines.

For each quasar, we retain the maximum absolute standardized residual across all epochs, yielding a distribution of peak deviations across the sample. Exceedances above $u$ are modeled by the Generalized Pareto Distribution (GPD) \citep{Coles2001}:
\begin{equation}
P(z > u + y \mid z > u) = \left(1 + \xi\,\frac{y}{\beta}\right)^{-1/\xi},
\end{equation}
where $\xi$ is the shape parameter and $\beta$ is the scale parameter. A positive $\xi$ indicates a heavy-tailed distribution, consistent with rare but extreme deviations expected from flare-like events. We set $u$ at the 95th percentile of the maximum $|z|$ distribution, a standard choice in POT applications that balances sufficient tail sample size for a reliable GPD fit against contamination from the bulk of the distribution \citep{Coles2001}, and fit the GPD using maximum likelihood estimation. The detection boundary is defined as the $z$-score at which the false alarm probability (FAP) equals 1\%. For $\xi \neq 0$, this is given by
\begin{equation}
z_{\mathrm{threshold}} = u + \frac{\beta}{\xi}\left[\left(\frac{N \cdot \mathrm{FAP}}{n_u}\right)^{-\xi} - 1\right],
\end{equation}
where $n_u$ is the number of exceedances above $u$ and $N$ is the total number of objects. Any spurious residuals arising from DRW model mismatch at either baseline are addressed by the downstream VLM verification stage, which filters false positives. We verify that the detection boundary is robust to the choice of $u$ by repeating the GPD fit across percentiles from the 90th to the 98th; the shape parameter and boundary remain stable across this range for both baselines (Appendix~\ref{app:evt_stability}).

\subsubsection{EVT on GRU Residuals}
\label{sec:evt_gru}

The standardized residuals from the GRU applied to the simulated validation set are well calibrated, with a mean of 0.06 and variance of 1.05 (ideal: 0 and 1 respectively). However, the distribution exhibits a kurtosis of 15.67 (compared to 0 for a Gaussian), confirming heavy-tailed departures from normality and motivating the use of EVT. After applying the trained GRU to an independent set of simulated DRW light curves (Section~\ref{sec:data}), we compute the standardized residual at each epoch as
\begin{equation}
z_i = \frac{m_i - \mu_i}{\sigma_i}.
\end{equation}
To correct for any residual bias in the model predictions, we calibrate the $z$-scores using the global mean $\bar{z}$ and standard deviation $s_z$ computed across all residuals from the validation set:
\begin{equation}
z_{\mathrm{cal},i} = \frac{z_i - \bar{z}}{s_z}.
\end{equation}
Applying the POT procedure to the distribution of maximum calibrated $|z|$ across the 9,258 simulated quasars yields a detection threshold of $8.69\sigma$. Any object in the Stripe~82 dataset whose maximum calibrated $z$-score exceeds this threshold is flagged as a candidate, producing 51 candidates (Figure~\ref{fig:evt_gru}).

\begin{figure*}[ht]
    \centering
    \includegraphics[width=\textwidth]{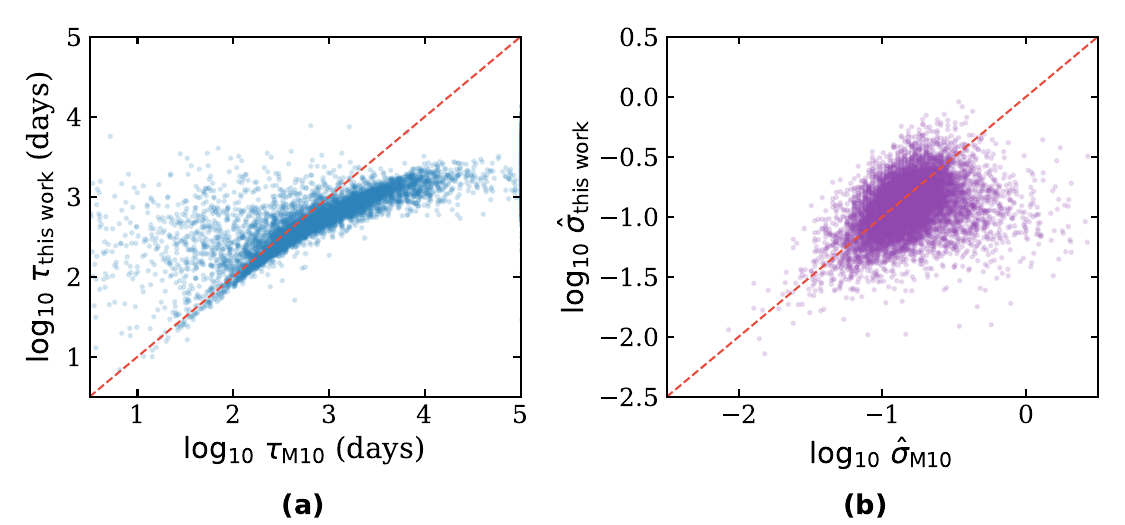}
    \caption{Comparison of $r$-band DRW parameters estimated in this work using the iterative OU process with those derived by \citet{2010ApJ...721.1014M} (M10). (a)~Damping timescale $\tau$ (Spearman $\rho = 0.86$), showing strong agreement along the 1:1 line (dashed). (b)~Variability amplitude $\hat{\sigma}$ ($\rho = 0.39$), showing larger scatter and a systematic offset above the 1:1 line, indicating that our estimates are generally higher.}
    \label{fig:macleod_comparison}
\end{figure*}

\subsubsection{EVT on Iterative OU Residuals}
For the iterative OU baseline, the standardized residuals (Equation~\ref{eq:ou_resid}) already incorporate both the OU conditional variance and photometric errors, so no additional calibration step is required. Applying the same POT procedure to the distribution of $\max_i |z_i|$ across the 9,104 converged quasars yields a detection threshold of $3.73\sigma$, producing 92 candidates (Figure~\ref{fig:evt_ou}).
The lower threshold relative to the GRU ($3.73\sigma$ vs.\ $8.69\sigma$) arises because the iterative OU process is fitted directly to the observed data, producing residuals that closely follow the expected OU statistics and yielding a tighter null distribution. In contrast, the GRU residuals are computed on simulated data, where model mismatch between the GRU predictions and the DRW realizations introduces additional scatter, broadening the null distribution and inflating the tail.

\subsection{Recognition Engine: VLM Benchmarking}

Once flare candidates are identified by the anomaly scoring stage, the final step is verification. In most flare detection pipelines, this requires manual visual inspection by a human expert, which becomes a bottleneck at scale. Traditional convolutional neural networks (CNNs) are not well suited for this task, as quasar light curve data are typically sparsely and irregularly sampled, requiring interpolation or binning that introduces artifacts and degrades temporal fidelity. Vision Language Models (VLMs), by contrast, operate directly on rendered light curve images, sidestepping the need for uniform temporal sampling. The recognition engine (Figure~\ref{fig:vlm_engine}) automates this verification step using VLMs.

\begin{table}[H]
\centering
\caption{VLMs benchmarked for the recognition engine.}
\label{tab:vlm_models}
\begin{tabular}{lc}
\hline
Model & Access \\
\hline
Qwen2.5VL-7b & Open-weight \\
Qwen2.5VL-7b-QLoRA & Fine-tuned \\
Qwen3VL-235B-A22 & Open-weight \\
Qwen3VL-30B & Open-weight \\
Qwen-3.5-plus & Proprietary \\
Mistral3Large & Open-weight \\
Claude 3 Haiku & Proprietary \\
Kimi-k2.5 & Open-weight \\
Grok-4.1-fast & Proprietary \\
GPT-5 & Proprietary \\
GPT-5-nano & Proprietary \\
GPT-5 mini & Proprietary \\
\hline
\end{tabular}
\end{table}

We benchmark 12 VLMs on the five-class classification task described in Section~\ref{sec:data}, spanning both open-weight and proprietary models (Table~\ref{tab:vlm_models}). All models are evaluated on the same test set of 4,630 light curves with 926 per class. Each model receives a light curve image paired with a structured prompt instructing it to classify the morphology as one of five classes (DRW, Spike, Gaussian, Gamma, FRED) and provide a brief shape description and confidence level. The same prompt format is used for all 12 VLMs to ensure a consistent evaluation.

To assess whether domain-specific fine-tuning improves performance, we perform parameter-efficient fine-tuning of Qwen2.5VL-7b using QLoRA \citep{Dettmers2023}. The model is quantized to 4-bit precision using NF4 quantization with double quantization. Low-rank adapters with rank $r = 8$ and scaling factor $\alpha = 16$ are applied to the query, key, value, and output projection layers of the attention mechanism. The model is fine-tuned for 1 epoch using the AdamW optimizer with a learning rate of $10^{-4}$ and a cosine learning rate scheduler on the training split described in Section~\ref{sec:data}. Benchmarking results and the selection of classifier and evaluator VLMs for the recognition engine are presented in Section~\ref{sec:results}.

\begin{figure*}[ht]
    \centering
    \includegraphics[width=0.7\textwidth]{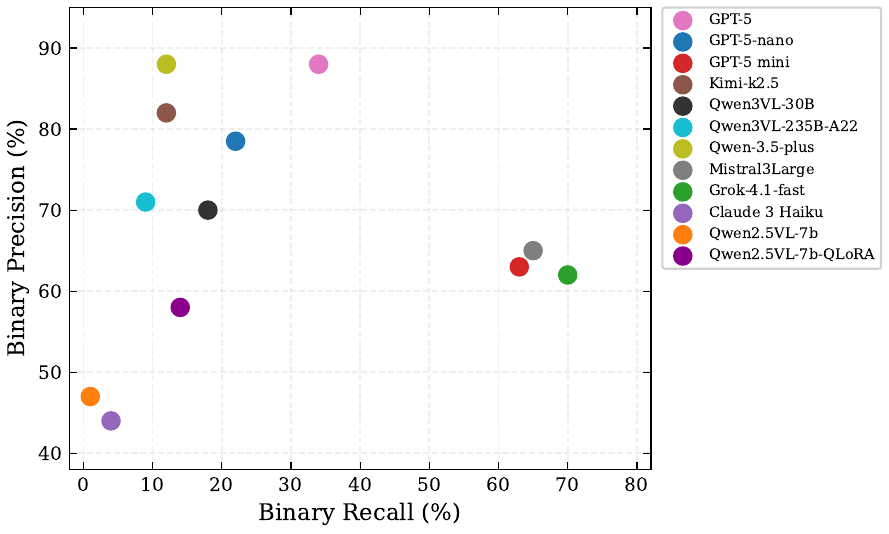}
    \caption{Binary flare detection precision versus recall for all 12 benchmarked VLMs, obtained by collapsing the five classes into flare (FRED, Gaussian, Gamma) and non-flare (DRW, Spike). Qwen-3.5-plus achieves the highest precision (${\sim}\,88\%$), while Grok-4.1-fast achieves the highest recall (${\sim}\,70\%$). Based on these results, we select Grok-4.1-fast as the high-recall classifier (Classifier~A), Qwen-3.5-plus as the high-precision classifier (Classifier~B), and GPT-5 as the evaluator for the recognition engine described in Section~\ref{sec:framework}.}
    \label{fig:precision_recall}
    \vspace{2mm}
\end{figure*}

\section{Results}
\label{sec:results}

\subsection{Iterative OU Baseline}

\citet{Suberlak2021} showed that the DRW parameters estimated by \citet{2010ApJ...721.1014M} can be improved by extending the light curve baseline, as short baselines relative to the characteristic timescale lead to biased parameter recovery. Motivated by this, we independently re-estimate the DRW parameters by fitting an OU process directly to the observed data using the iterative masking procedure described in Section~\ref{sec:iter_ou}. The procedure converges for 9,104 of the 9,258 quasars, with a median of 1 iteration and a median of 0 masked epochs, indicating that the vast majority of light curves are well described by the DRW model without requiring outlier removal. Figure~\ref{fig:macleod_comparison} compares our re-estimated DRW parameters with those of \citet{2010ApJ...721.1014M}. The damping timescales show strong agreement (Spearman $\rho = 0.86$), while the variability amplitudes show a weaker correlation ($\rho = 0.39$), consistent with the finding of \citet{Suberlak2021} that $\hat{\sigma}$ is more sensitive to baseline length and data cleaning procedures than $\tau$.

\subsection{VLM Benchmarking}

The five-class classification accuracy for all 12 benchmarked VLMs is shown in Appendix~\ref{app:vlm_accuracy}. GPT-5 achieves the highest accuracy at 42.8\%, followed by GPT-5-nano and GPT-5 mini at 32.6\%. All models exceed the 20\% random baseline for five classes, but none achieve reliable fine-grained morphological discrimination, suggesting that five-class light curve classification from images alone remains a challenging task for current VLMs.

The base Qwen2.5VL-7b model predicts non-flare for the vast majority of inputs, achieving 97\% accuracy on the non-flare class but failing to detect most flare morphologies. Fine-tuning with QLoRA substantially improves the model's ability to identify flare events: the fraction of flare light curves correctly classified as a flare type (rather than non-flare) increases from ${\sim}\,3\%$ to ${\sim}\,15\%$. However, fine-tuning introduces a systematic bias toward the Gamma class, with most flare predictions collapsing into this single category regardless of the true morphology. Confusion matrices for both the base and fine-tuned models are presented in Appendix~\ref{app:confusion_matrices}.

Since the recognition engine ultimately requires a binary flare/non-flare decision, we collapse the five classes into flare (FRED, Gaussian, Gamma) and non-flare (DRW, Spike) and evaluate binary precision and recall. Figure~\ref{fig:precision_recall} shows the results for all models. GPT-5 and Qwen-3.5-plus achieve the highest binary precision (${\sim}\,88\%$), while Grok-4.1-fast achieves the highest recall (${\sim}\,70\%$). Based on the dual-classifier design described in Section~\ref{sec:framework}, we select Grok-4.1-fast as the high-recall classifier, Qwen-3.5-plus as the high-precision classifier, and GPT-5 as the evaluator due to its highest overall accuracy and balance of both precision and recall.

\subsection{Confirmed Flares}
\begin{figure}[H]
    \centering
    \includegraphics[width=0.8\columnwidth]{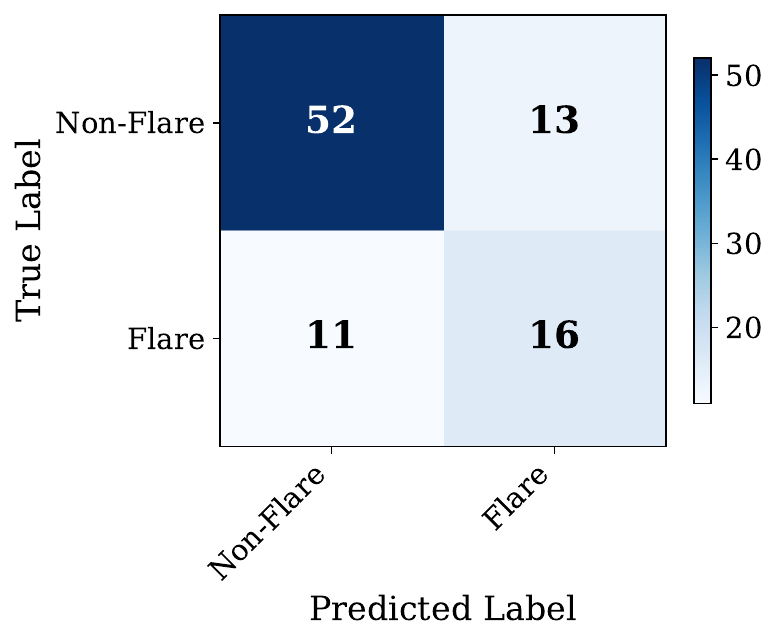}
    \caption{Confusion matrix for the recognition engine applied to the 92 flare candidates from the iterative OU baseline. True labels are determined by independent human verification of both $r$-band and $g$-band light curves. The recognition engine achieves a precision of 55.2\% and a recall of 59.3\%, with 16 true positives, 13 false positives, 11 false negatives, and 52 true negatives.}
    \label{fig:cm_re_ou}
\end{figure}

\vspace{-1mm}
Applying the iterative OU baseline to the 9,104 converged quasars and computing the maximum absolute standardized residual for each object, the EVT procedure yields a detection threshold of $3.73\sigma$, producing 92 candidates. The recognition engine classifies 29 of these as flares. Independent human verification of all 92 candidates identifies 27 as genuine flares based on visual inspection of both $r$-band and $g$-band light curves. Figure~\ref{fig:cm_re_ou} presents the resulting confusion matrix: of the 29 objects flagged by the recognition engine, 16 are confirmed by human inspection (true positives) and 13 are not (false positives), while 11 human-verified flares are missed by the recognition engine (false negatives), yielding a precision of 55.2\% and a recall of 59.3\%.

\begin{figure}[H]
    \centering
    \includegraphics[width=0.8\columnwidth]{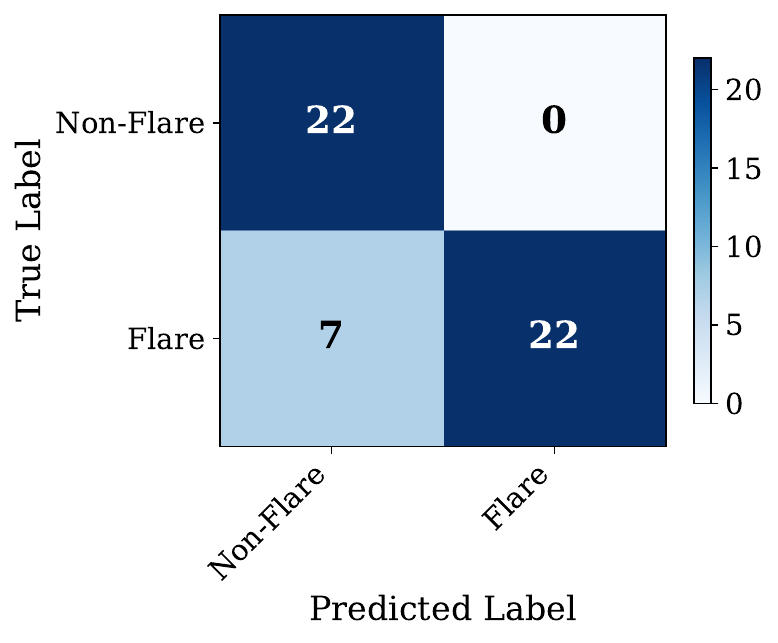}
    \caption{Confusion matrix for the recognition engine applied to the 51 flare candidates from the GRU baseline. True labels are determined by independent human verification of both $r$-band and $g$-band light curves. The recognition engine achieves a precision of 100\% and a recall of 75.9\%, with 22 true positives, zero false positives, 7 false negatives, and 22 true negatives.}
    \label{fig:cm_re_gru}
\end{figure}

For the GRU baseline, applying the trained model to the 9,258 Stripe~82 light curves and computing calibrated standardized residuals, the EVT procedure yields a detection threshold of $8.69\sigma$, producing 51 candidates. The recognition engine classifies 22 of these as flares. Human verification of all 51 candidates identifies 29 as genuine flares. Figure~\ref{fig:cm_re_gru} presents the confusion matrix: all 22 objects flagged by the recognition engine are confirmed by human inspection (precision of 100\%), while 7 human-verified flares are missed (recall of 75.9\%).

Taking the union of human-verified flares from both baselines, we identify a total of 51 unique quasars exhibiting distinct flaring activity. Of these, 5 are detected by both baselines, 22 are detected exclusively by the iterative OU process, and 24 exclusively by the GRU. This complementarity confirms the motivation for deploying two baselines with distinct sensitivity profiles: the iterative OU process, fitted directly to the observed data, captures subtler deviations with a lower detection threshold, while the GRU, trained on simulated DRW light curves, enforces a more rigid baseline and is sensitive to transients that a flexible model might absorb into its fit. The light curves of all 51 confirmed flaring quasars, cross-checked against $g$-band data, are presented in the Appendix~\ref{app:flares}.

\section{Discussion}
\label{sec:discussions}

The FLARE framework identifies 51 flaring quasars from 9,258 objects in the SDSS Stripe~82 dataset, corresponding to a flare rate of ${\sim}\,0.55\%$. For comparison, \citet{10.1093/mnras/stx1456} found 51 flares from over 900,000 quasars in CRTS (${\sim}\,0.006\%$) using sigma-clipping on de-trended light curves. \citet{10.1093/mnras/stae721} applied Gaussian Process analysis to 9,035 ZTF Type~1 AGN light curves and identified 27 flare candidates (${\sim}\,0.3\%$) with a false-positive rate below 7\%. \citet{He_2025} conducted the largest systematic search to date using Bayesian blocks combined with Gaussian Processes on ZTF~DR23, constructing a coarse catalog of 28,504 flares and a refined catalog of 1,984 high-confidence flares. \citet{10.1093/mnras/stae1036} identified 11 quasars with flare/eclipse-like variability from ${\sim}\,83{,}000$ SDSS quasars using ZTF data (${\sim}\,0.013\%$). Our flare rate is comparable to that of \citet{10.1093/mnras/stae721} and significantly higher than the CRTS and ZTF-based studies, likely reflecting the deeper temporal baseline of Stripe~82 (${\sim}\,10$ years with ${\sim}\,60$--$80$ epochs per object) combined with the sensitivity of the EVT-based threshold, which is calibrated to the specific noise properties of the dataset rather than relying on a fixed sigma cut. A direct comparison of flare rates across surveys is difficult, however, as differences in cadence, photometric depth, baseline length, and detection methodology all influence the number of candidates recovered. Notably, FLARE is the first systematic flare search applied to the Stripe~82 quasar sample.
Of the 51 confirmed flares, only 5 are detected by both the GRU and iterative OU baselines, with 22 unique to the iterative OU process and 24 unique to the GRU. This limited overlap is not merely a consequence of the differing EVT thresholds ($8.69\sigma$ for the GRU versus $3.73\sigma$ for the iterative OU); it reflects fundamental differences in how the two baselines compute residuals. The GRU residuals are derived from a model trained on simulated DRW light curves and calibrated against a validation set (Section~\ref{sec:evt_gru}), whereas the iterative OU residuals are computed from a process fitted directly to the observed data with iterative outlier masking (Section~\ref{sec:iter_ou}). These distinct sensitivity profiles---the GRU enforcing a rigid, simulation-derived baseline and the iterative OU adapting to the real data---mean that each baseline preferentially detects different classes of deviations. The small overlap confirms that neither baseline alone provides a complete flare census and validates the dual-baseline design.
For the recognition engine, performance depends on the specific VLMs selected and the system prompt provided. While different prompting strategies may yield slightly different candidate counts, we do not expect the results to vary substantially provided the prompt correctly defines the classification task. Moreover, the dual-classifier and evaluator architecture (Section~\ref{sec:framework}) mitigates the impact of individual model biases, and the independent human verification provides a prompt-independent assessment of the final catalog. We note that the GRU baseline recognition engine achieves 100\% precision but 75.9\% recall, while the iterative OU baseline recognition engine achieves 55.2\% precision and 59.3\% recall. The lower precision on the OU candidates likely reflects the larger and more heterogeneous candidate pool (92 versus 51), which includes subtler deviations that are more difficult for VLMs to classify.

Several additional limitations of the current implementation should be noted. First, the recognition engine relies on proprietary VLMs (GPT-5, Grok-4.1-fast, Qwen-3.5-plus), which limits reproducibility. As open-weight VLMs continue to improve, future implementations may achieve comparable performance with fully reproducible models. Second, the variance regularization term in the GRU loss function acts as a soft constraint rather than achieving exact agreement with the OU predicted variance, suggesting that alternative variance calibration strategies may further improve the baseline model. Third, the human verification of candidates remains a manual step. The recognition engine substantially narrows the human-in-the-loop gap by automating the flare/non-flare classification that would otherwise require expert visual inspection, and this gap is expected to shrink further as VLM classification capabilities continue to improve.

The FLARE framework is designed to be generalizable beyond Stripe~82. The three-stage structure---baseline modeling, anomaly scoring, and recognition engine---is modular: each component can be replaced independently as better methods become available. For example, the physics-informed GRU could be substituted with Gaussian Processes for surveys with denser cadence, or the EVT threshold could be recalibrated for different noise regimes. The framework is directly applicable to ongoing and upcoming surveys such as ZTF and the Vera C.\ Rubin Observatory's Legacy Survey of Space and Time (LSST), which will monitor millions of quasars with higher cadence and photometric precision. 
As VLM capabilities continue to advance, the recognition engine is expected to become increasingly reliable, reducing dependence on manual verification. Physical characterization of the 51 identified flares---including analysis of their redshift distribution, black hole mass dependence, and possible physical origins---is deferred to a follow-up study.

\section{Conclusions}
\label{sec:conclusion}

We have presented FLARE, a modular three-stage framework for detecting transient flares in quasar light curves, and applied it to the SDSS Stripe~82 dataset of 9,258 spectroscopically confirmed quasars. Our main conclusions are as follows.

\begin{enumerate}

\item Two complementary baselines---a physics-informed probabilistic GRU trained on simulated DRW light curves with Ornstein--Uhlenbeck regularization, and an iterative OU process fitted directly to observed data with outlier masking---provide well-calibrated residuals suitable for statistical anomaly detection. The GRU residuals exhibit heavy tails (kurtosis $= 15.67$), motivating the use of Extreme Value Theory over fixed sigma thresholds.

\item A Peaks-Over-Threshold EVT analysis yields detection boundaries of $8.69\sigma$ (GRU baseline) and $3.73\sigma$ (iterative OU baseline) at 1\% false alarm probability, both stable across threshold percentiles from the 90th to the 98th. These produce 51 and 92 flare candidates, respectively.

\item Of the 51 confirmed flaring quasars, only 5 are detected by both baselines, with 22 unique to the iterative OU process and 24 unique to the GRU. This limited overlap validates the dual-baseline design: the two approaches have complementary sensitivity profiles and neither alone provides a complete flare census.

\item We benchmark 12 Vision Language Models on a five-class light curve classification task. While fine-grained morphological discrimination remains challenging (best five-class accuracy: 42.8\% by GPT-5), binary flare/non-flare classification achieves operationally useful precision and recall, enabling the dual-classifier recognition engine design. The recognition engine achieves 100\% precision with 75.9\% recall on the GRU candidates, and 55.2\% precision with 59.3\% recall on the iterative OU candidates.

\item Parameter-efficient fine-tuning of Qwen2.5VL-7b via QLoRA improves flare sensitivity from ${\sim}\,3\%$ to ${\sim}\,15\%$ but introduces systematic class bias, indicating that domain-specific fine-tuning of VLMs for astronomical classification remains an open problem.

\item The FLARE framework is modular and generalizable: each stage---baseline modeling, anomaly scoring, and recognition engine---can be independently replaced as improved methods become available, making it directly applicable to current and future surveys such as ZTF and LSST.

\end{enumerate}

\section*{Acknowledgments}

We acknowledge and thank MITACS for their Globalink Research Internship program, during which the author had fruitful early discussions that laid the foundation for this work. We also acknowledge the developers of the Vision Language Models used in this work: OpenAI---GPT-5, GPT-5-nano, and GPT-5 mini (August 2025); xAI---Grok-4.1-fast (November 2025); Alibaba Cloud---Qwen2.5-VL-7b-Instruct (August 2024), Qwen3-VL-235B-A22B-Instruct (February 2026), Qwen3-VL-30B-A3B (October 2025), and Qwen-3.5-plus (February 2026); Mistral AI---Mistral Large 3 (December 2025); Anthropic---Claude 3 Haiku (March 2024); and Moonshot AI---Kimi-k2.5 (January 2026).

This work made use of the quasar light curve data and DRW parameters from \citet{2010ApJ...721.1014M}.

Funding for the Sloan Digital Sky Survey~V has been provided by the Alfred P.\ Sloan Foundation, the Heising-Simons Foundation, the National Science Foundation, and the Participating Institutions. SDSS acknowledges support and resources from the Center for High-Performance Computing at the University of Utah. SDSS telescopes are located at Apache Point Observatory, funded by the Astrophysical Research Consortium and operated by New Mexico State University, and at Las Campanas Observatory, operated by the Carnegie Institution for Science. The SDSS web site is \url{www.sdss.org}.

SDSS is managed by the Astrophysical Research Consortium for the Participating Institutions of the SDSS Collaboration, including the Carnegie Institution for Science, Chilean National Time Allocation Committee (CNTAC) ratified researchers, Caltech, the Gotham Participation Group, Harvard University, Heidelberg University, The Flatiron Institute, The Johns Hopkins University, L'\'{E}cole polytechnique f\'{e}d\'{e}rale de Lausanne (EPFL), Leibniz-Institut f\"{u}r Astrophysik Potsdam (AIP), Max-Planck-Institut f\"{u}r Astronomie (MPIA Heidelberg), Max-Planck-Institut f\"{u}r Extraterrestrische Physik (MPE), Nanjing University, National Astronomical Observatories of China (NAOC), New Mexico State University, The Ohio State University, Pennsylvania State University, Smithsonian Astrophysical Observatory, Space Telescope Science Institute (STScI), the Stellar Astrophysics Participation Group, Universidad Nacional Aut\'{o}noma de M\'{e}xico, University of Arizona, University of Colorado Boulder, University of Illinois at Urbana-Champaign, University of Toronto, University of Utah, University of Virginia, Yale University, and Yunnan University.

\noindent\textit{Software:} NumPy \citep{numpy},
SciPy \citep{scipy},
PyTorch \citep{pytorch},
celerite \citep{celerite},
eztao \citep{eztao},
pyextremes (\url{https://github.com/georgebv/pyextremes}),
Hugging Face Transformers \citep{huggingface},
PEFT (\url{https://github.com/huggingface/peft}),
TRL (\url{https://github.com/huggingface/trl}),
Matplotlib \citep{matplotlib},
Pandas \citep{pandas}

\noindent\textit{Data and Code:} The code and simulated data used in this work are publicly available at this \href{https://github.com/zerozole/FLARE}{link}.

\bibliographystyle{apsrev4-1}

% You should give the same name for your .bbl as your main .tex
% since it is a requirement for posting on ArXiv.
\bibliography{oja_template}
\begin{appendix}
\section{EVT Threshold Stability Plots}
\label{app:evt_stability}

\begin{figure}[H]

    \centering

    \includegraphics[width=0.8\columnwidth]{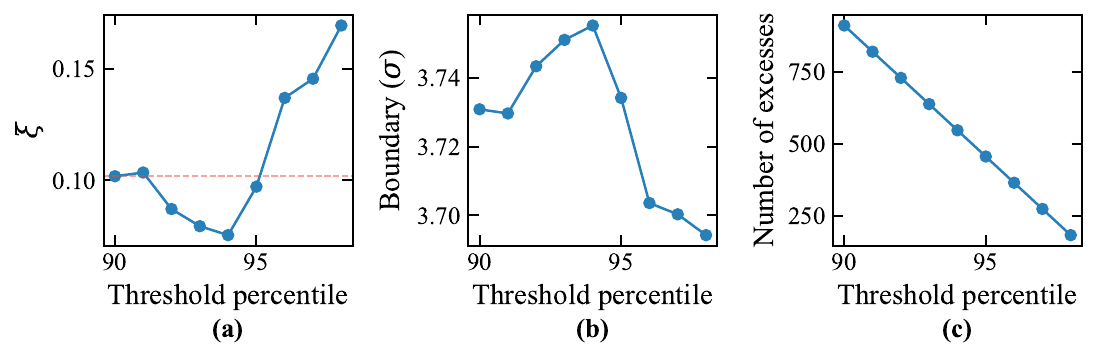}

    \caption{Stability of the GPD fit for the iterative OU baseline across threshold percentiles from the 90th to the 98th. (a)~The shape parameter $\xi$ remains stable around ${\sim}\,0.10$ (dashed red line). (b)~The detection boundary varies by less than $0.06\sigma$ across percentiles. (c)~The number of tail exceedances decreases linearly with increasing percentile, as expected.}

    \label{fig:evt_stability_ou}

\end{figure}

\begin{figure}[H]

    \centering

    \includegraphics[width=0.8\columnwidth]{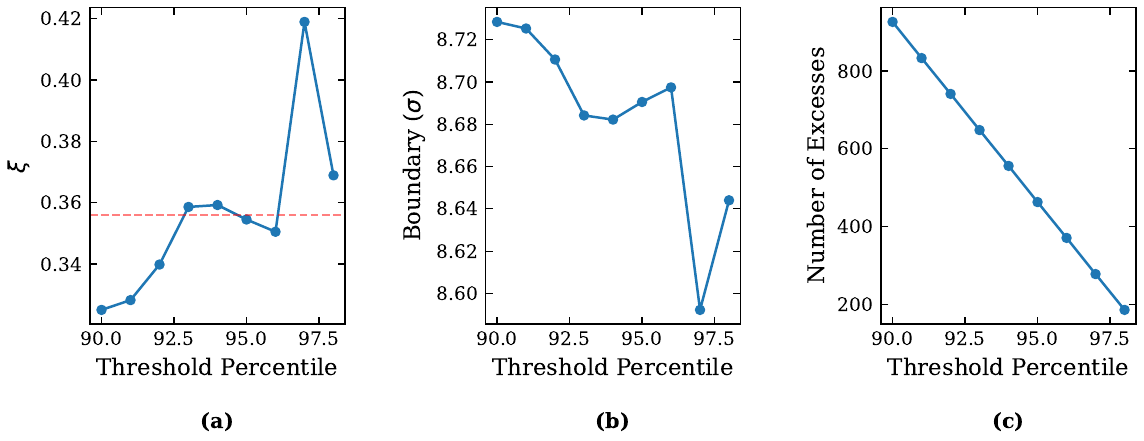}

    \caption{Stability of the GPD fit for the GRU baseline across threshold percentiles from the 90th to the 98th. (a)~The shape parameter $\xi$ remains stable around ${\sim}\,0.35$ (dashed red line). (b)~The detection boundary varies by less than $0.15\sigma$ across percentiles. (c)~The number of tail exceedances decreases linearly with increasing percentile, as expected.}

    \label{fig:evt_stability_gru}

\end{figure}
\vspace{-10mm}
\section{VLM Accuracy Plot}
\label{app:vlm_accuracy}
\begin{figure}[H]

    \centering

    \includegraphics[width=0.7\columnwidth]{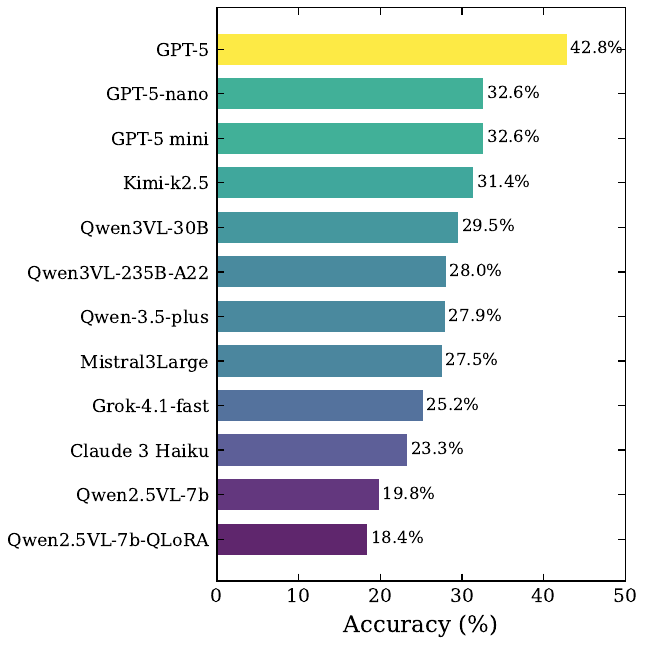}

    \caption{Five-class classification accuracy for all 12 benchmarked VLMs on the test set of 4,630 light curves (926 per class). GPT-5 achieves the highest accuracy at 42.8\%. All models exceed the 20\% random baseline.}

    \label{fig:acc_plot}

\end{figure}

\section{Qwen2.5VL-7b finetuned and base model confusion matrices}
\vspace{-5mm}
\label{app:confusion_matrices}
\begin{figure}[H]

    \centering

    \includegraphics[width=0.6\columnwidth]{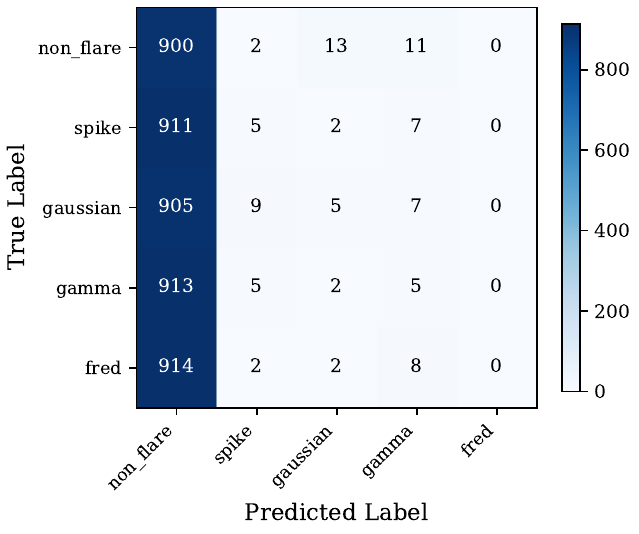}

    \caption{Confusion matrix for the base Qwen2.5VL-7b model on the five-class test set (926 light curves per class). The model predicts non-flare for the vast majority of inputs across all classes, achieving 97\% accuracy on the non-flare class but correctly identifying only ${\sim}\,3\%$ of flare light curves.}

    \label{fig:cm_base}

\end{figure}
\begin{figure}[H]

    \centering

    \includegraphics[width=0.6\columnwidth]{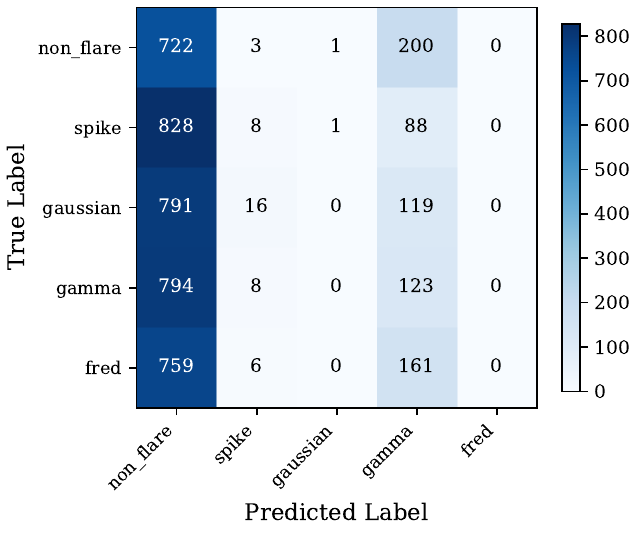}

    \caption{Confusion matrix for the QLoRA fine-tuned Qwen2.5VL-7b model on the five-class test set (926 light curves per class). Fine-tuning improves flare detection---the fraction of flare light curves correctly classified as a flare type increases from ${\sim}\,3\%$ to ${\sim}\,15\%$---but introduces a systematic bias toward the Gamma class, with most flare predictions collapsing into this single category regardless of the true morphology.}

    \label{fig:cm_qlora}

\end{figure}

\section{Light curves containing flares}
\label{app:flares}
\vspace{-5mm}
\begin{figure}[H]
      \centering
      \includegraphics[width=0.48\columnwidth]{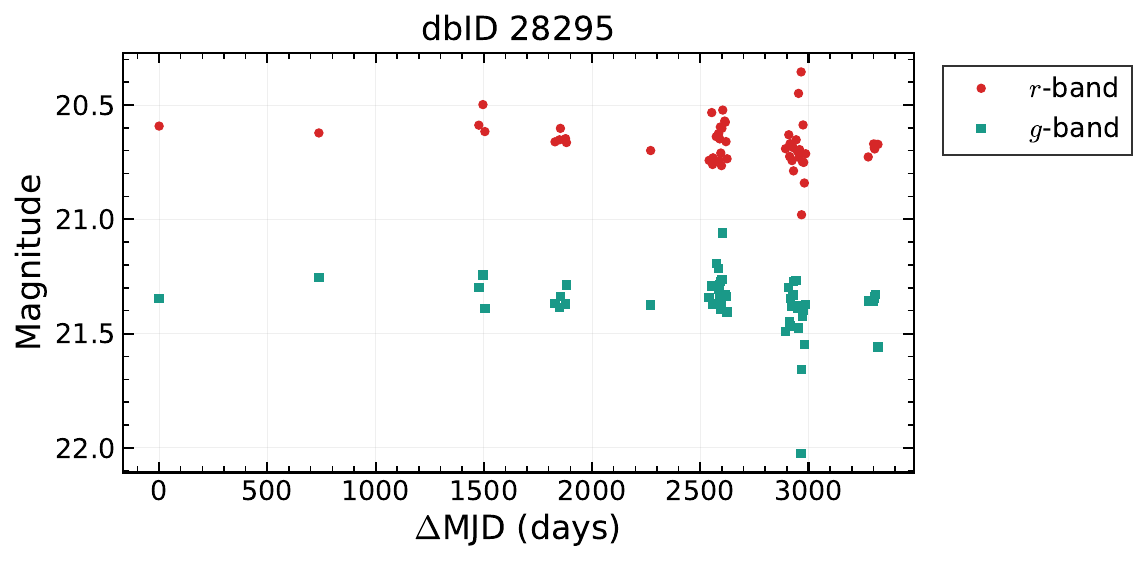}
      \hfill
      \includegraphics[width=0.48\columnwidth]{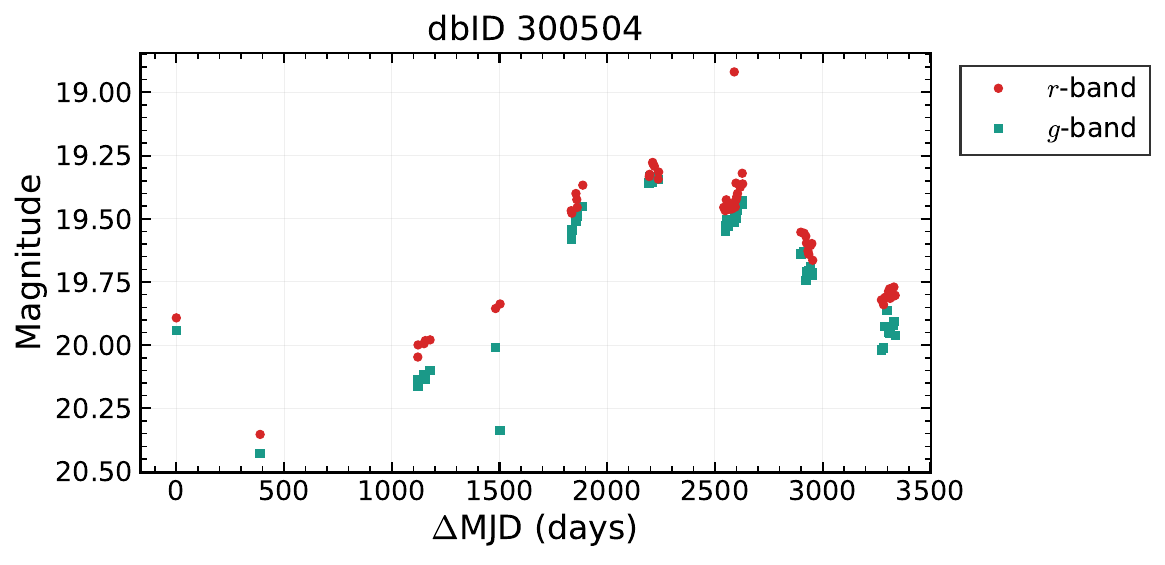}
      \caption{$r$-band (red) and $g$-band (green) light curves for the 51 confirmed flaring quasars identified by the FLARE framework. Each light curve is titled with the quasar's database identifier (dbID) from the \citet{2010ApJ...721.1014M} catalog. The horizontal axis shows the time elapsed since the first observation in Modified Julian Days ($\Delta$MJD), and the vertical axis shows the apparent magnitude. The correlated variability in both bands confirms that the detected flares are astrophysical in origin and not instrumental artifacts.}
      \label{fig:flare_gallery}
  \end{figure}

  \begin{figure}[H]\ContinuedFloat
      \centering
      \includegraphics[width=0.48\columnwidth]{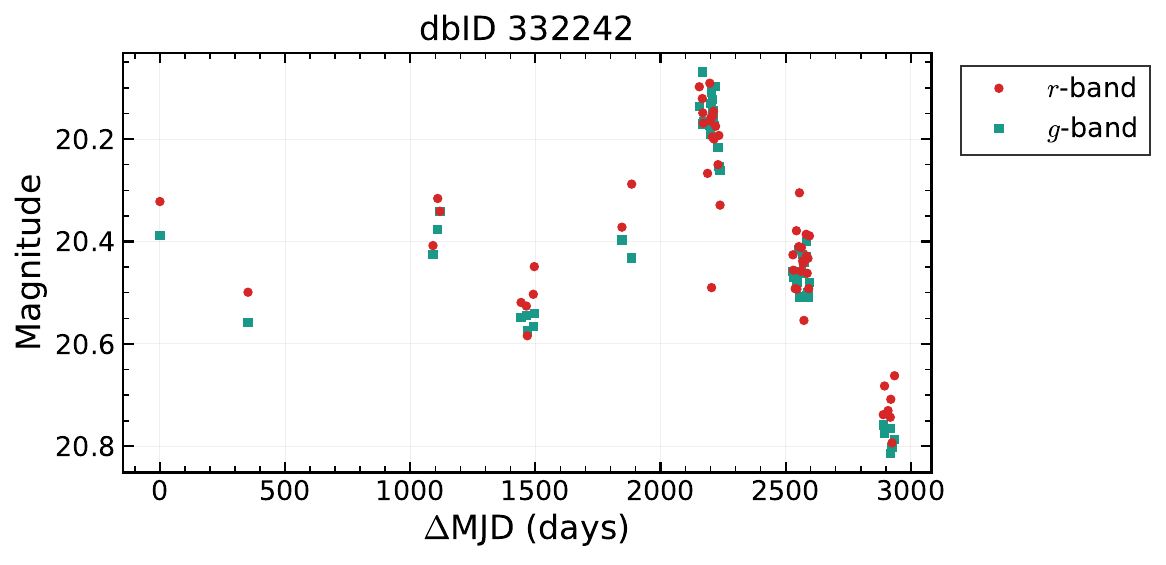}
      \hfill
      \includegraphics[width=0.48\columnwidth]{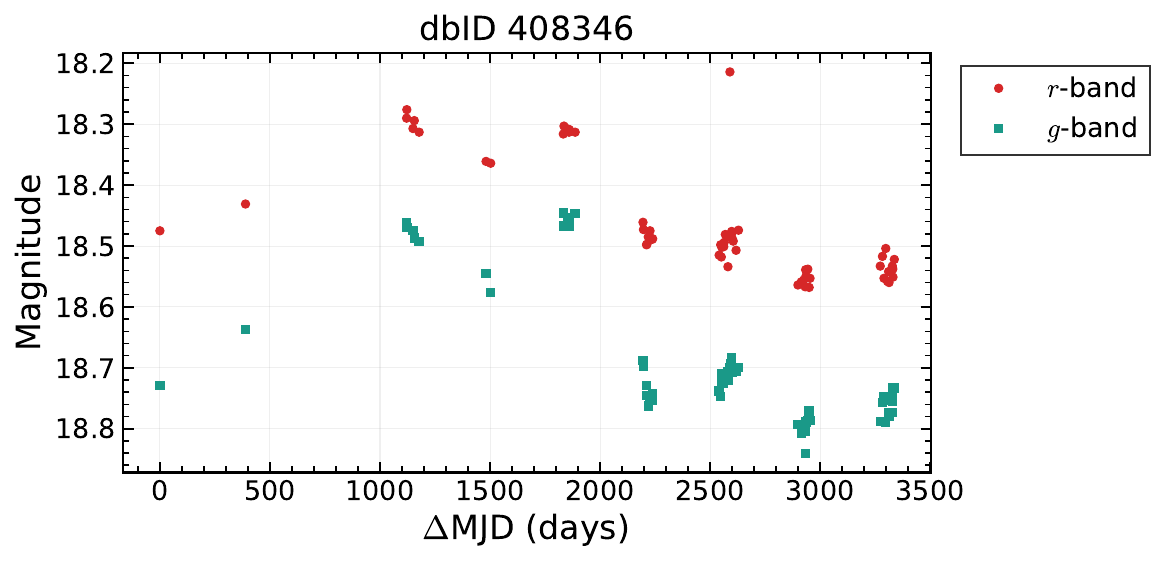}
      \caption{continued.}
  \end{figure}

  \begin{figure}[H]\ContinuedFloat
      \centering
      \includegraphics[width=0.48\columnwidth]{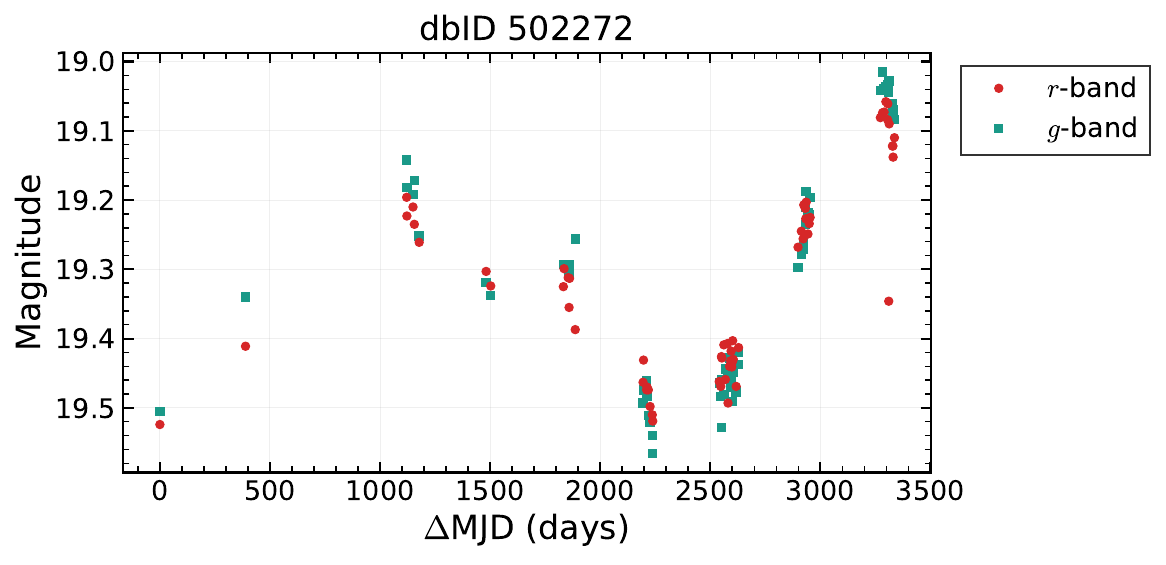}
      \hfill
      \includegraphics[width=0.48\columnwidth]{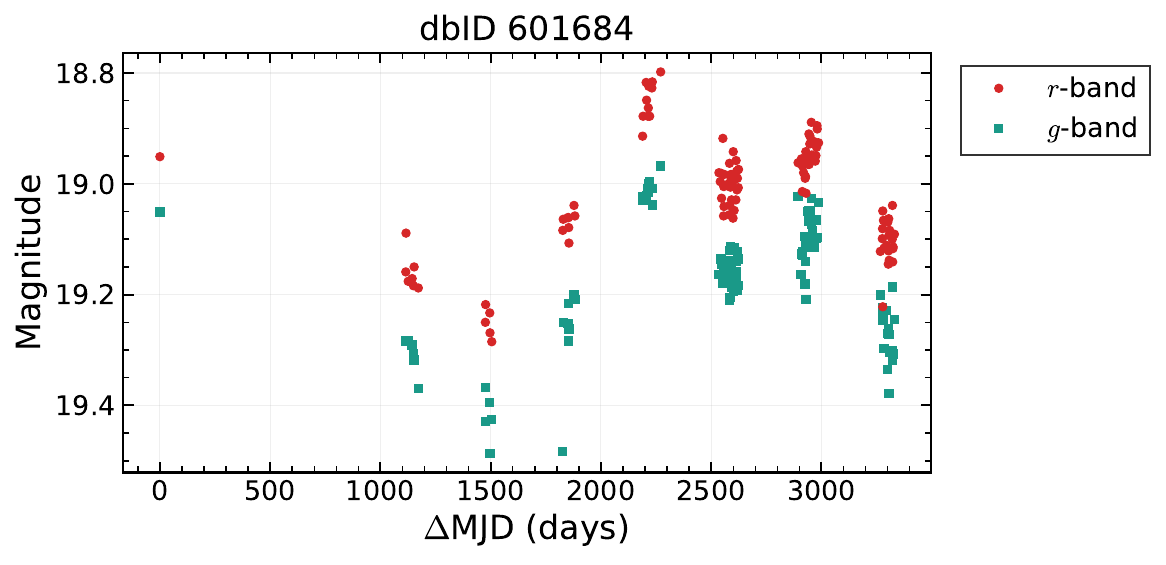}
      \caption{continued.}
  \end{figure}

  \begin{figure}[H]\ContinuedFloat
      \centering
      \includegraphics[width=0.48\columnwidth]{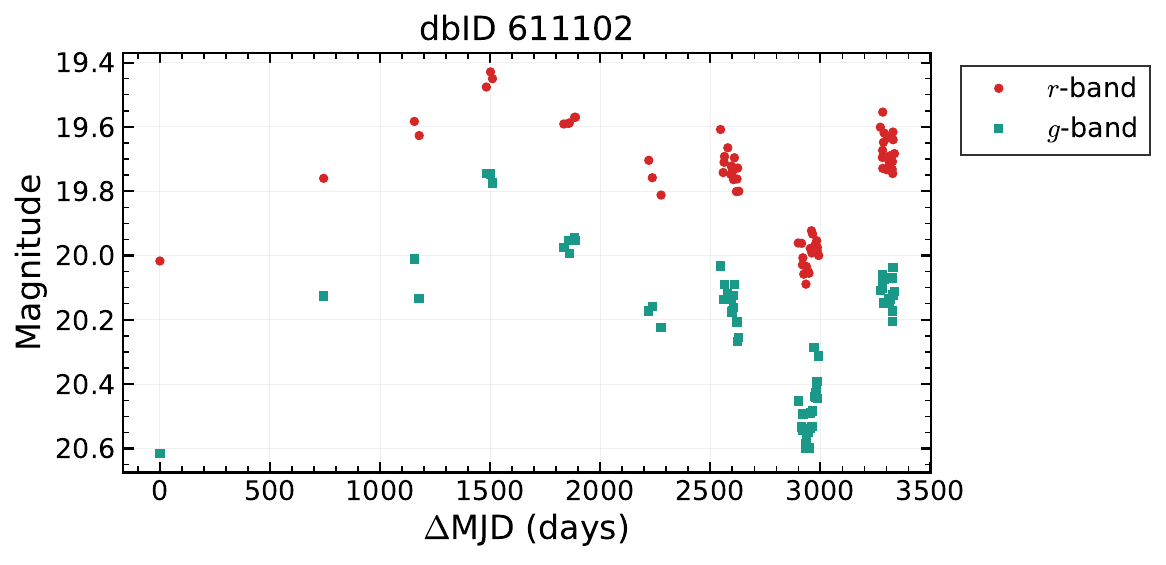}
      \hfill
      \includegraphics[width=0.48\columnwidth]{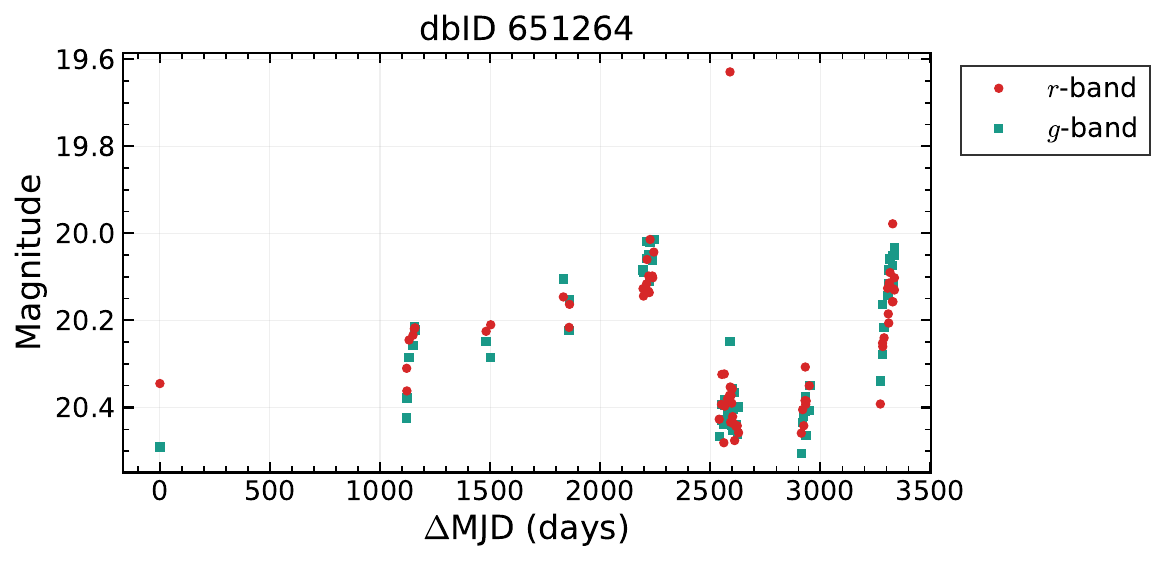}
      \caption{continued.}
  \end{figure}

  \begin{figure}[H]\ContinuedFloat
      \centering
      \includegraphics[width=0.48\columnwidth]{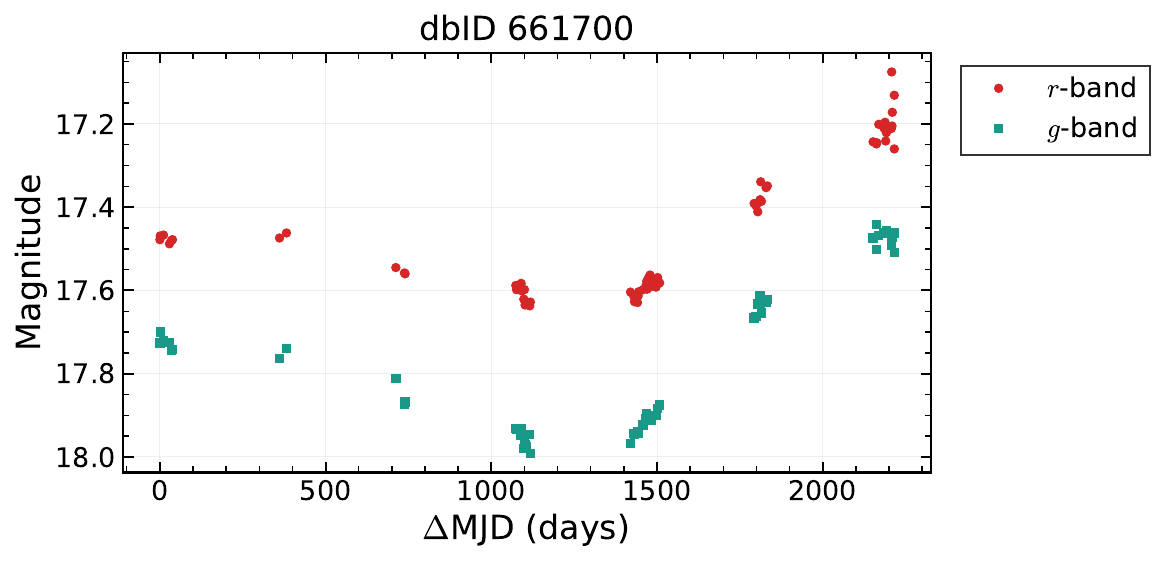}
      \hfill
      \includegraphics[width=0.48\columnwidth]{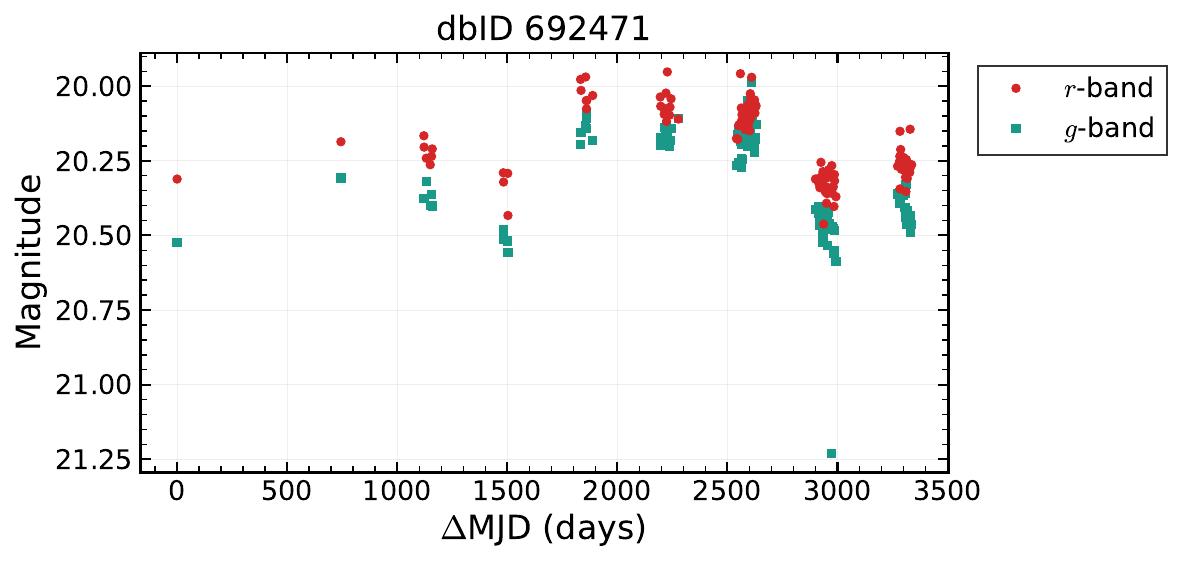}
      \caption{continued.}
  \end{figure}

  \begin{figure}[H]\ContinuedFloat
      \centering
      \includegraphics[width=0.48\columnwidth]{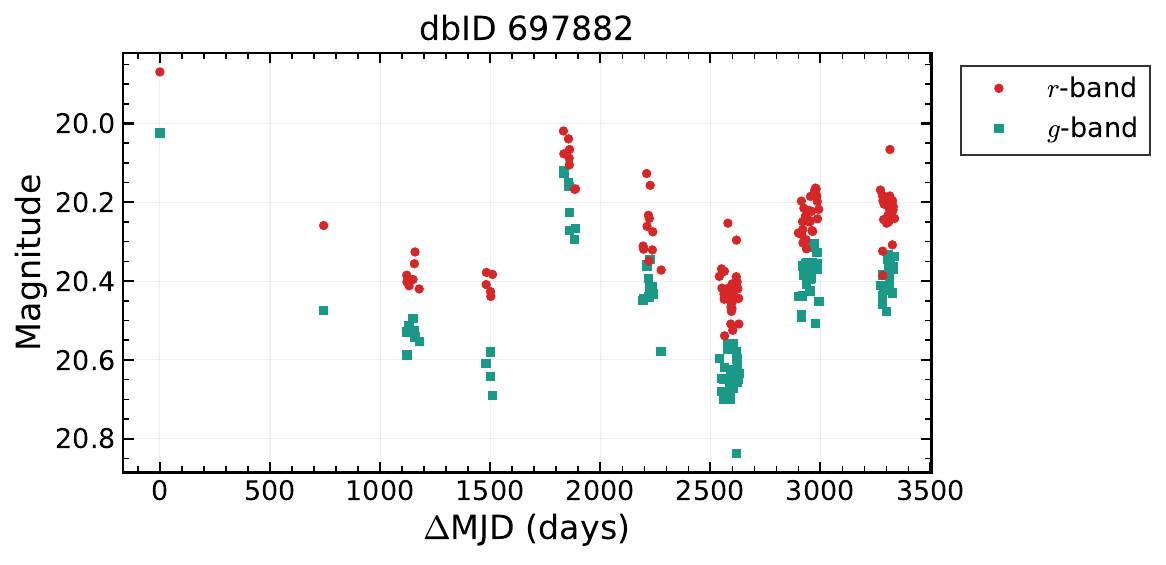}
      \hfill
      \includegraphics[width=0.48\columnwidth]{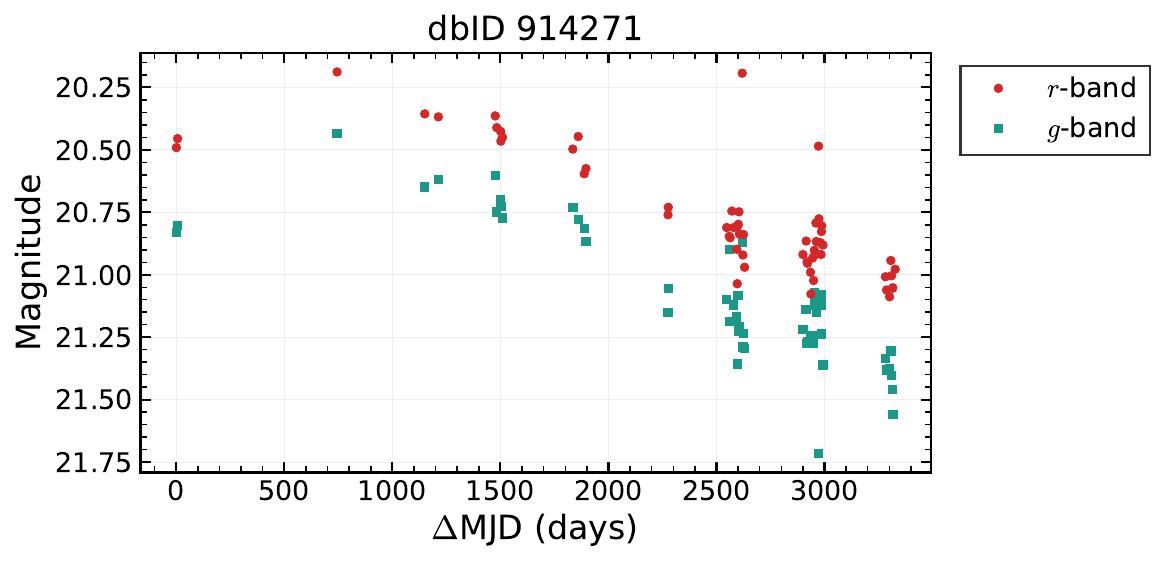}
      \caption{continued.}
  \end{figure}

  \begin{figure}[H]\ContinuedFloat
      \centering
      \includegraphics[width=0.48\columnwidth]{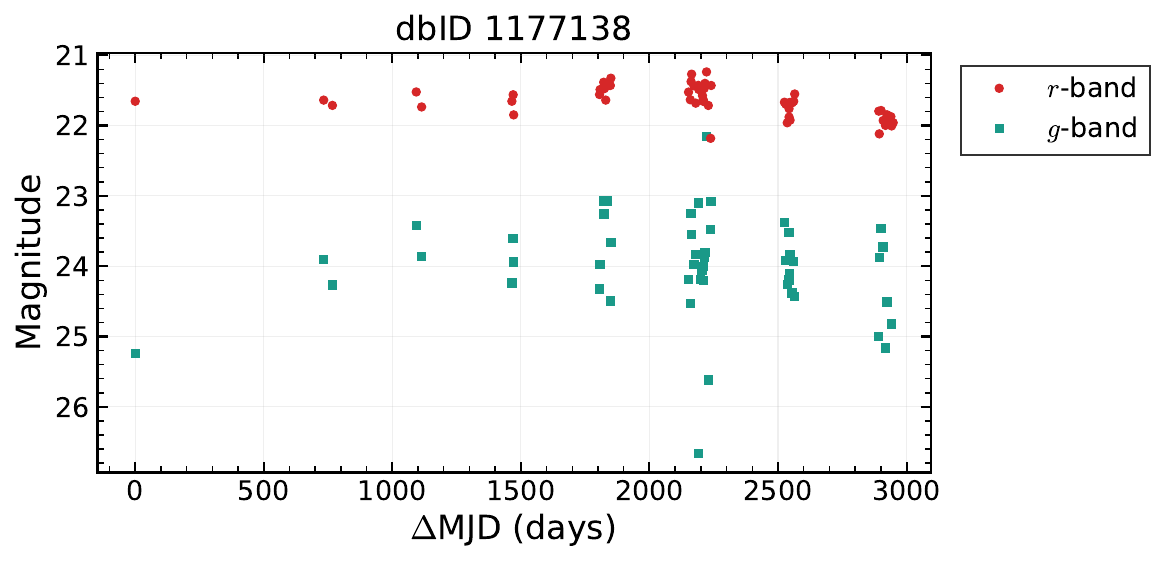}
      \hfill
      \includegraphics[width=0.48\columnwidth]{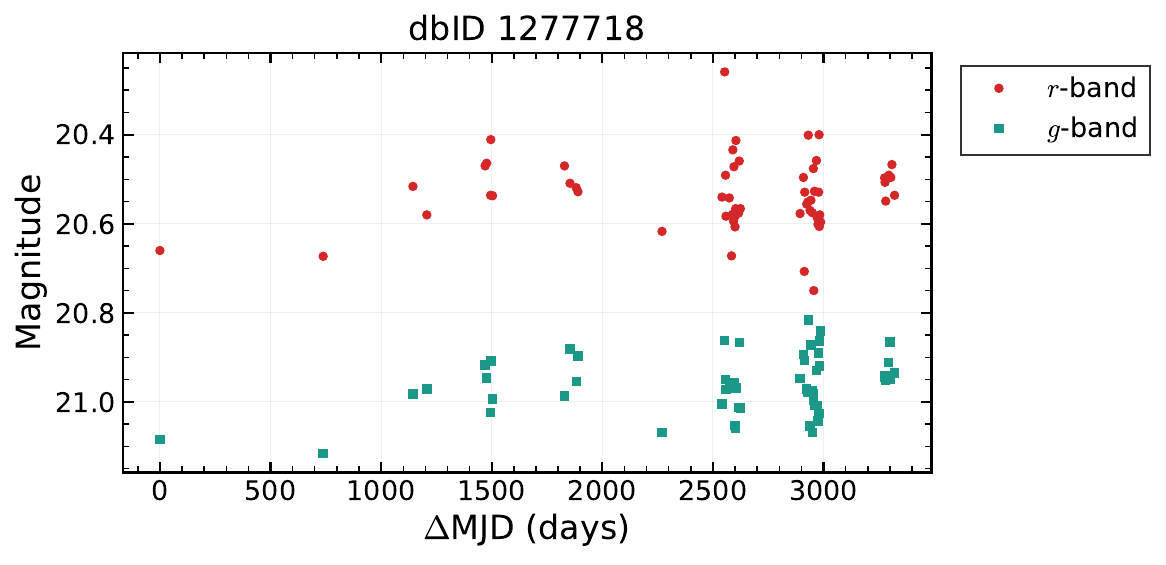}
      \caption{continued.}
  \end{figure}

  \begin{figure}[H]\ContinuedFloat
      \centering
      \includegraphics[width=0.48\columnwidth]{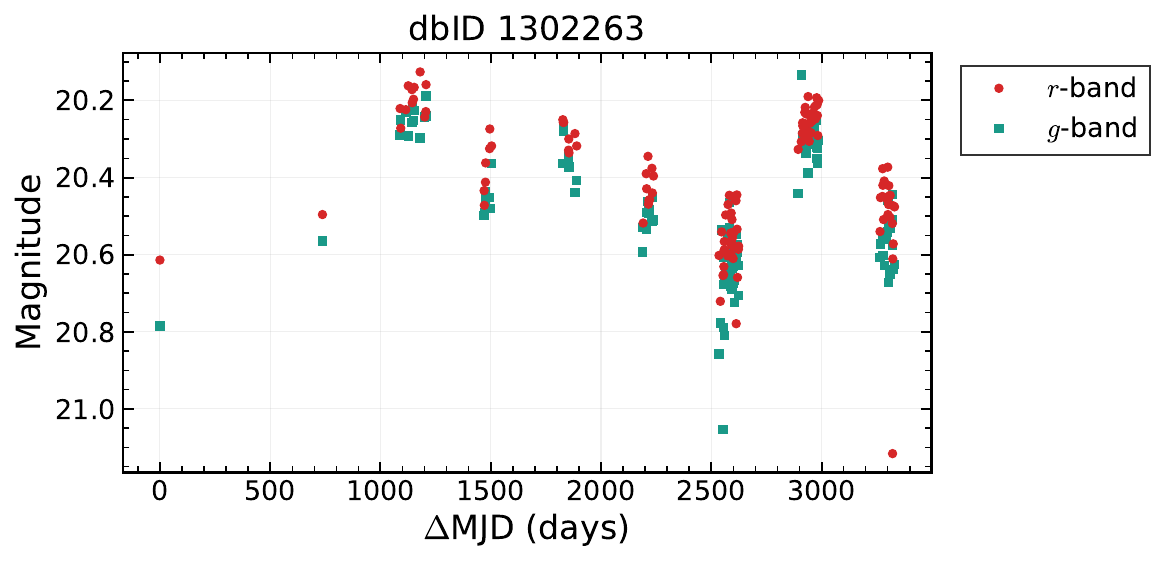}
      \hfill
      \includegraphics[width=0.48\columnwidth]{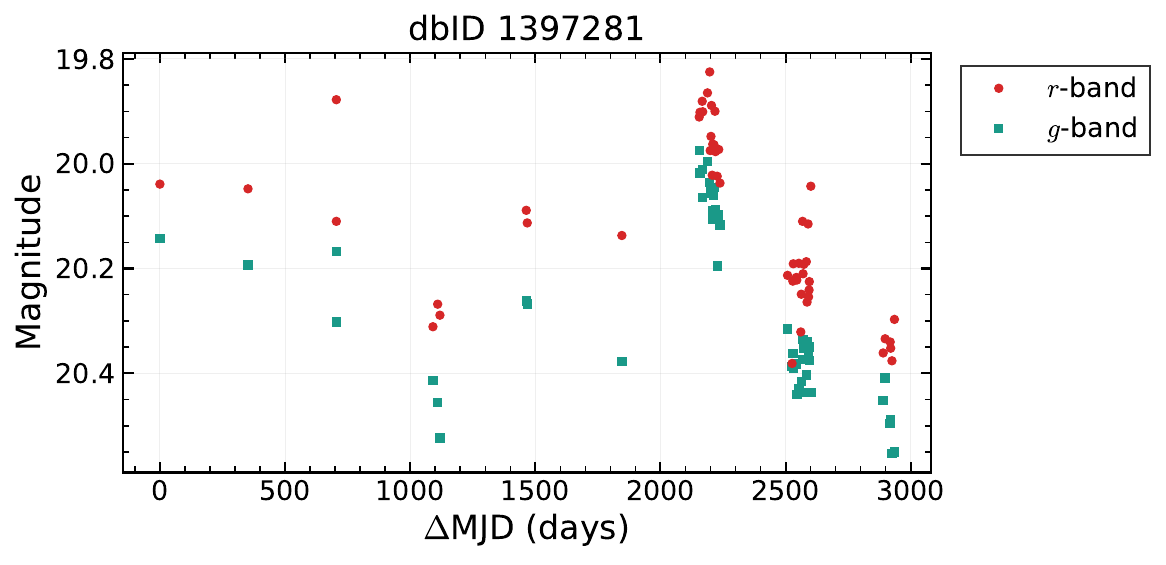}
      \caption{continued.}
  \end{figure}

  \begin{figure}[H]\ContinuedFloat
      \centering
      \includegraphics[width=0.48\columnwidth]{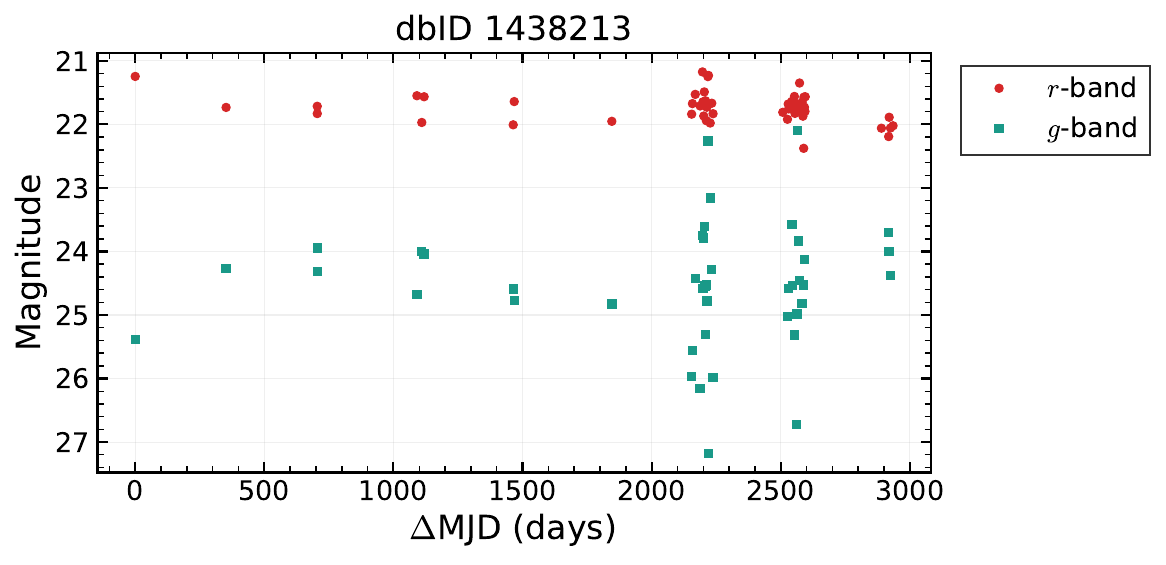}
      \hfill
      \includegraphics[width=0.48\columnwidth]{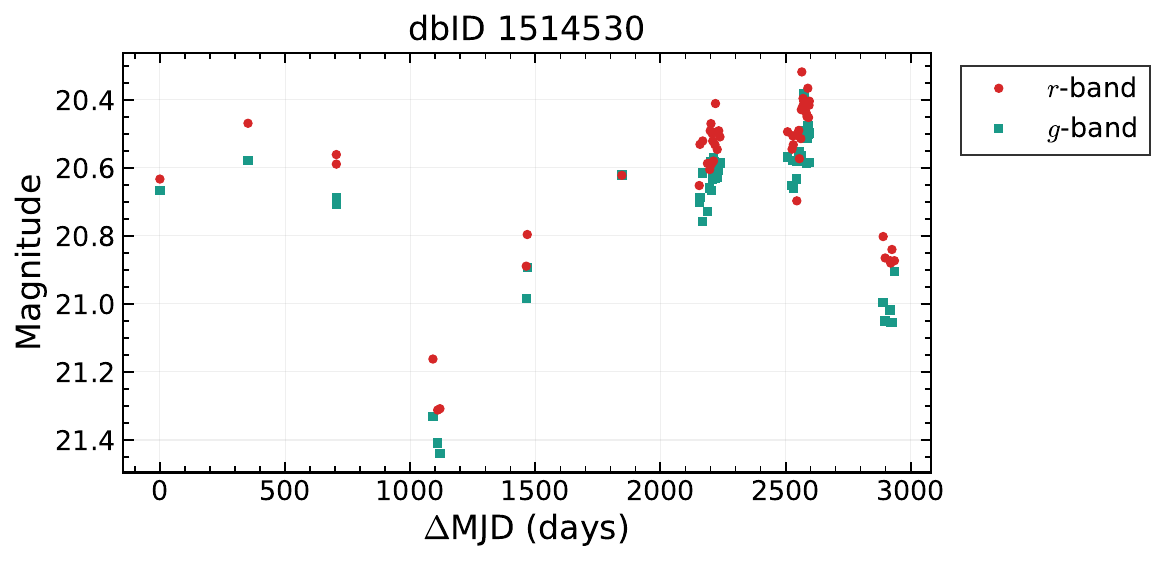}
      \caption{continued.}
  \end{figure}

  \begin{figure}[H]\ContinuedFloat
      \centering
      \includegraphics[width=0.48\columnwidth]{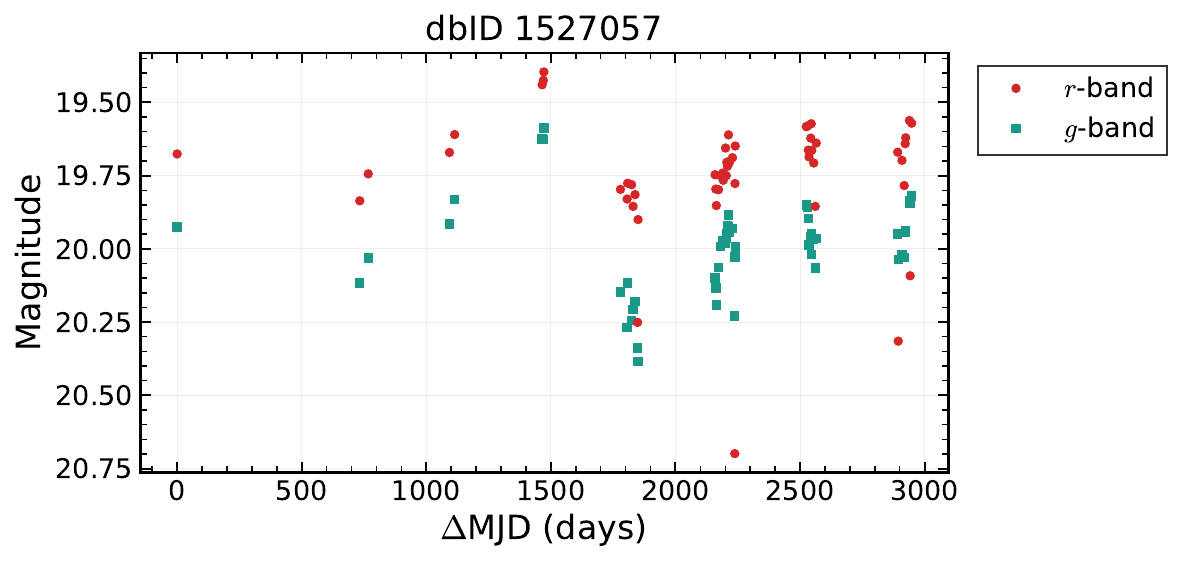}
      \hfill
      \includegraphics[width=0.48\columnwidth]{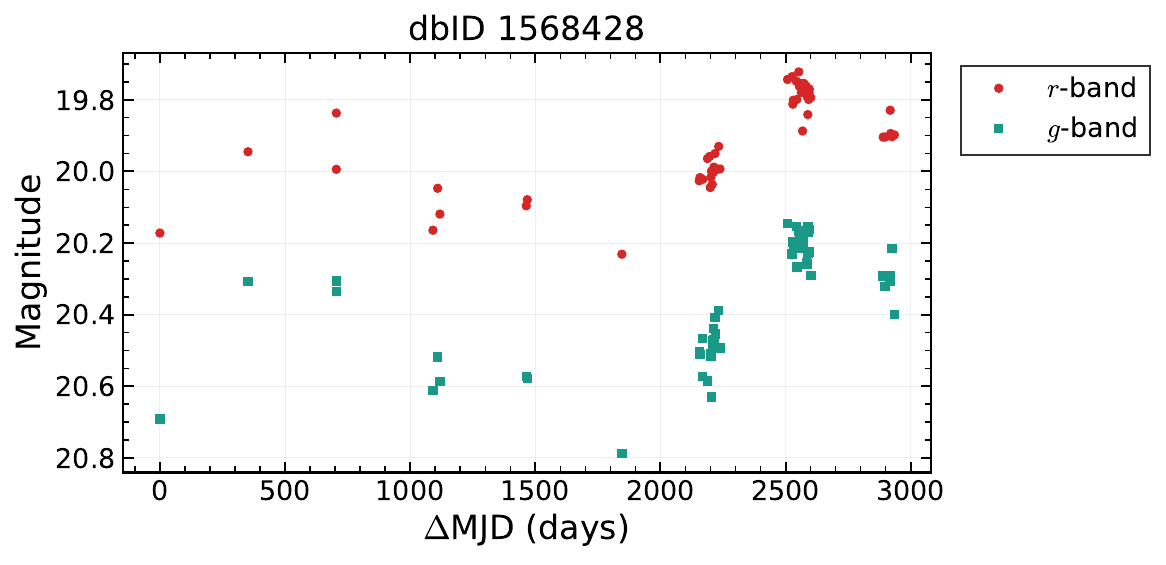}
      \caption{continued.}
  \end{figure}

  \begin{figure}[H]\ContinuedFloat
      \centering
      \includegraphics[width=0.48\columnwidth]{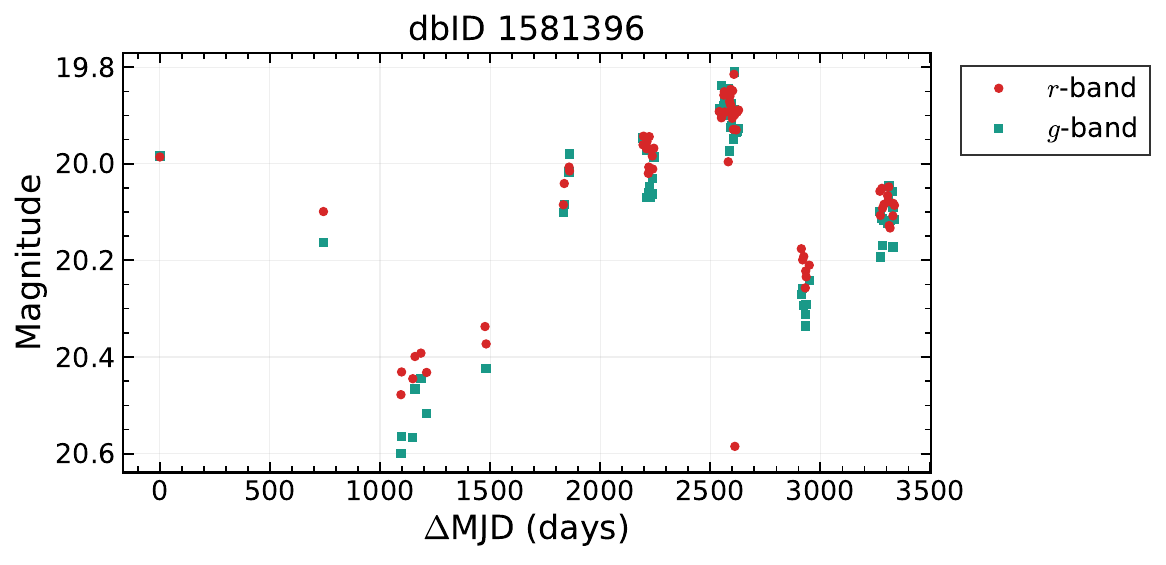}
      \hfill
      \includegraphics[width=0.48\columnwidth]{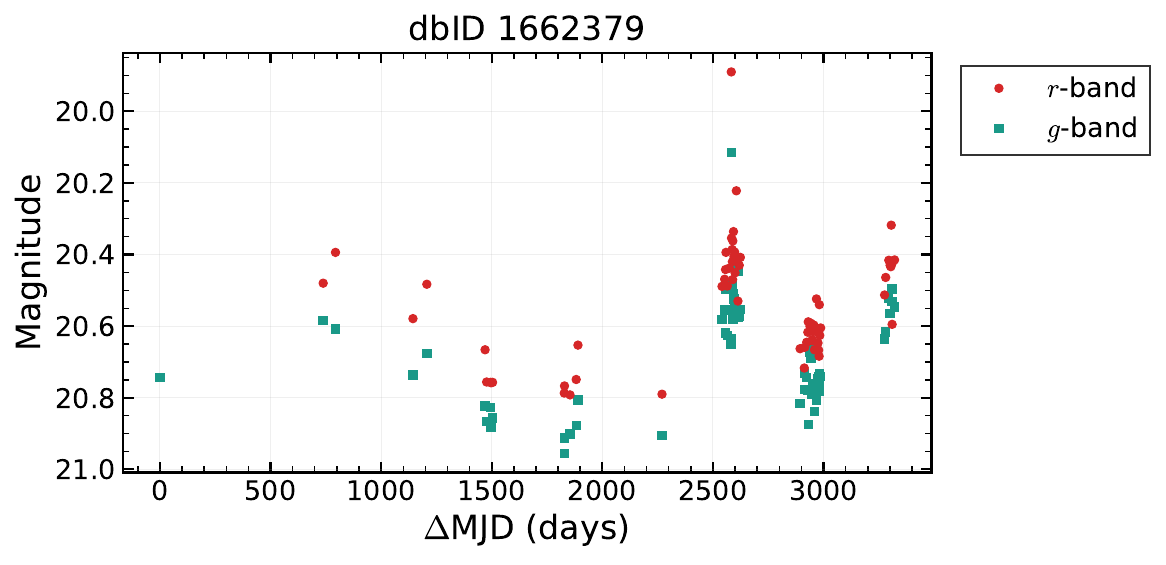}
      \caption{continued.}
  \end{figure}

  \begin{figure}[H]\ContinuedFloat
      \centering
      \includegraphics[width=0.48\columnwidth]{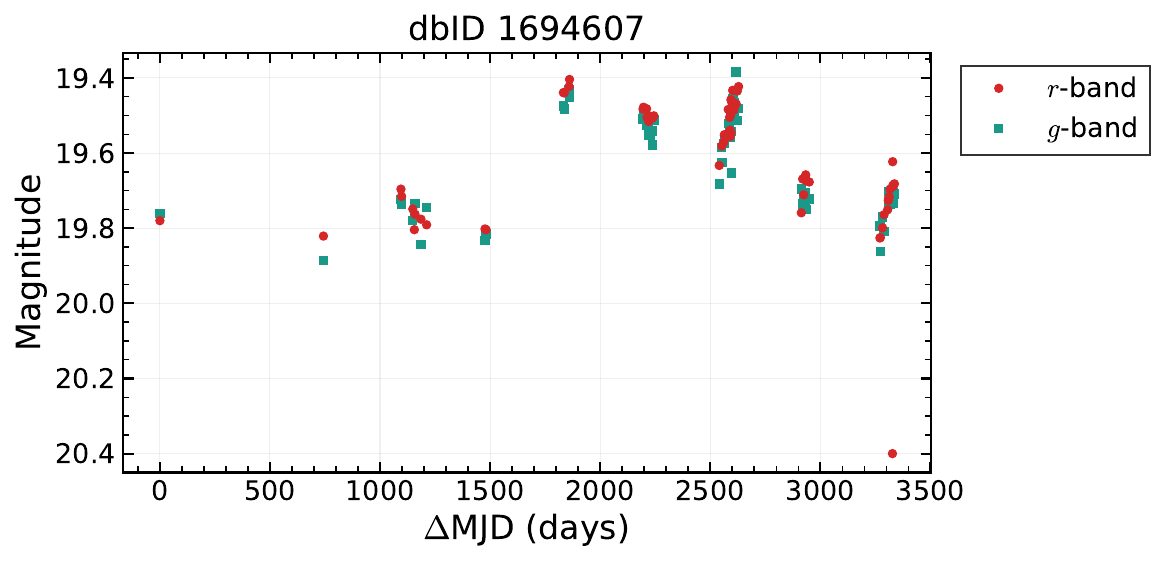}
      \hfill
      \includegraphics[width=0.48\columnwidth]{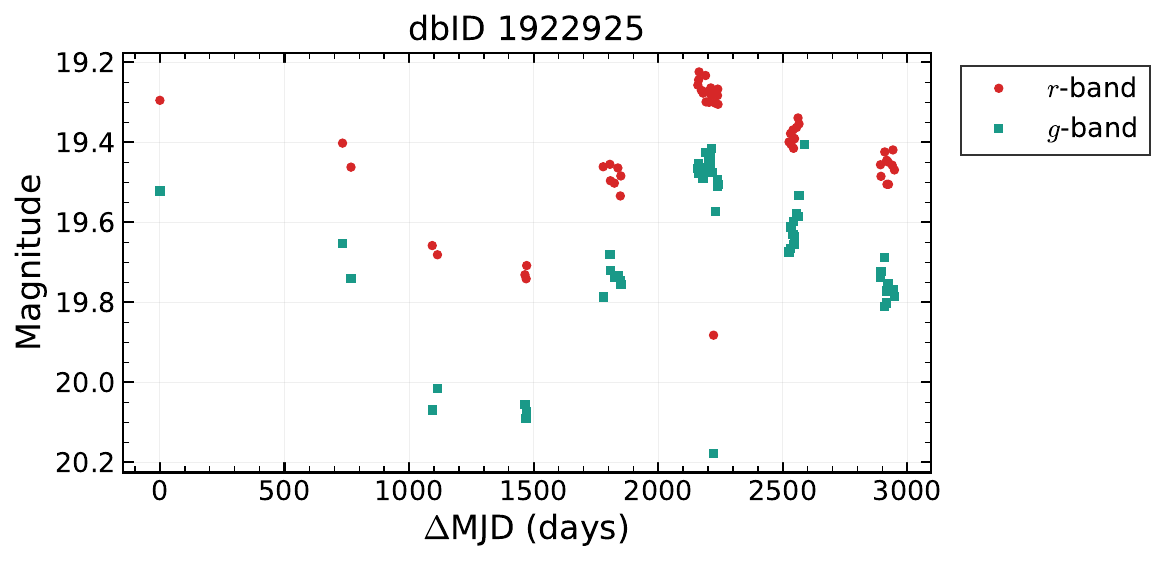}
      \caption{continued.}
  \end{figure}

  \begin{figure}[H]\ContinuedFloat
      \centering
      \includegraphics[width=0.48\columnwidth]{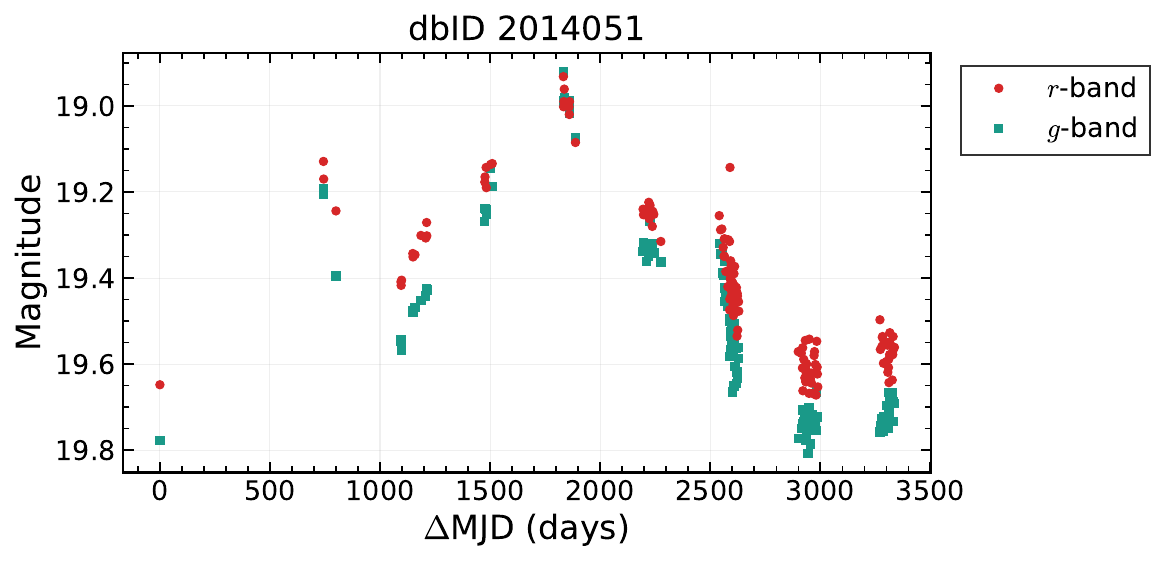}
      \hfill
      \includegraphics[width=0.48\columnwidth]{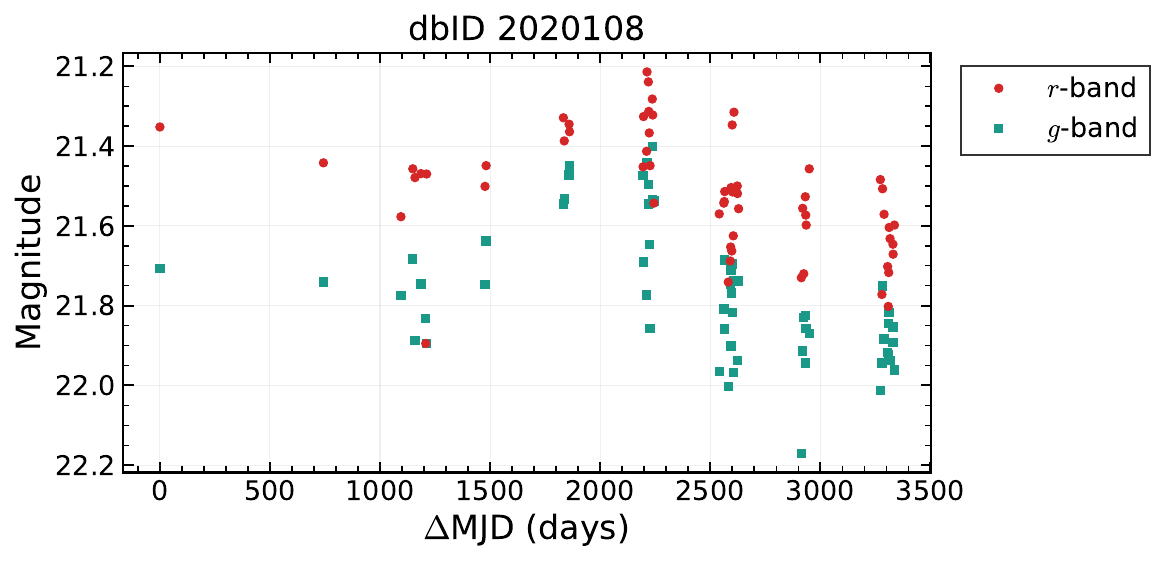}
      \caption{continued.}
  \end{figure}

  \begin{figure}[H]\ContinuedFloat
      \centering
      \includegraphics[width=0.48\columnwidth]{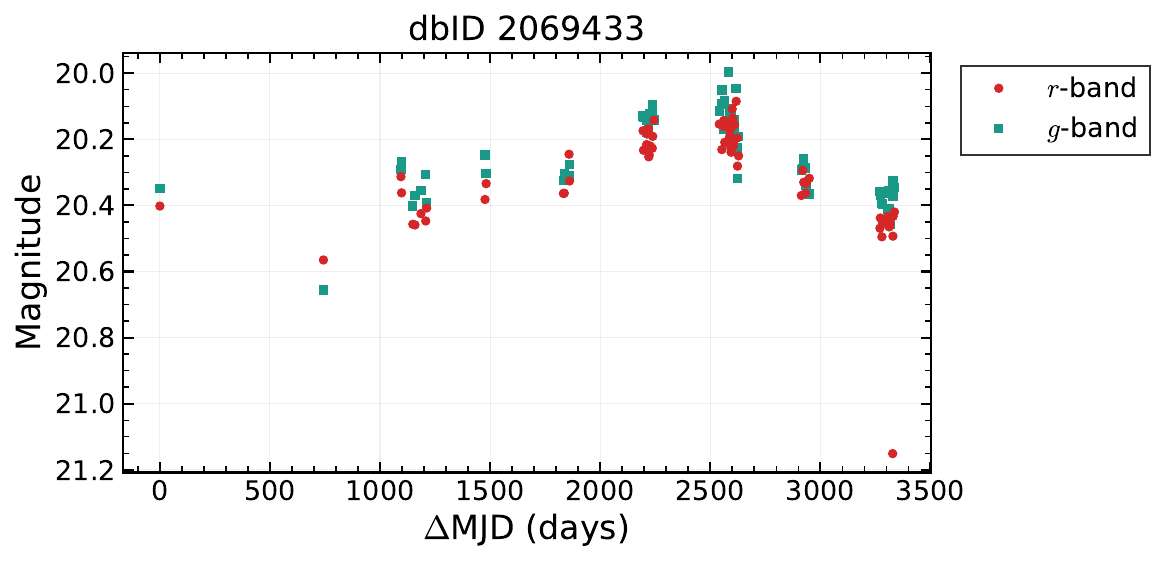}
      \hfill
      \includegraphics[width=0.48\columnwidth]{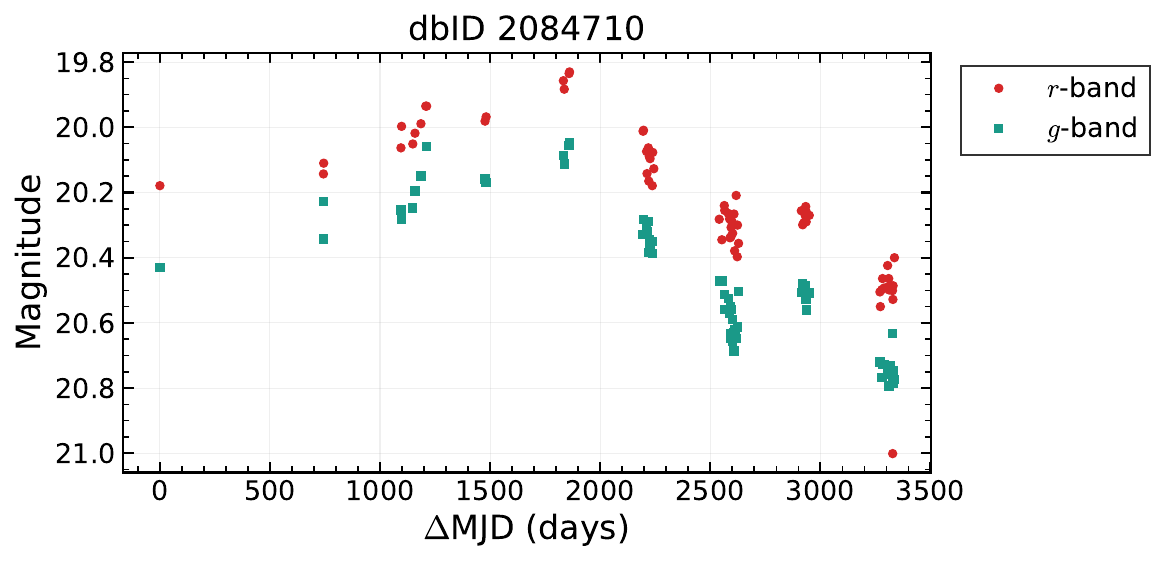}
      \caption{continued.}
  \end{figure}

  \begin{figure}[H]\ContinuedFloat
      \centering
      \includegraphics[width=0.48\columnwidth]{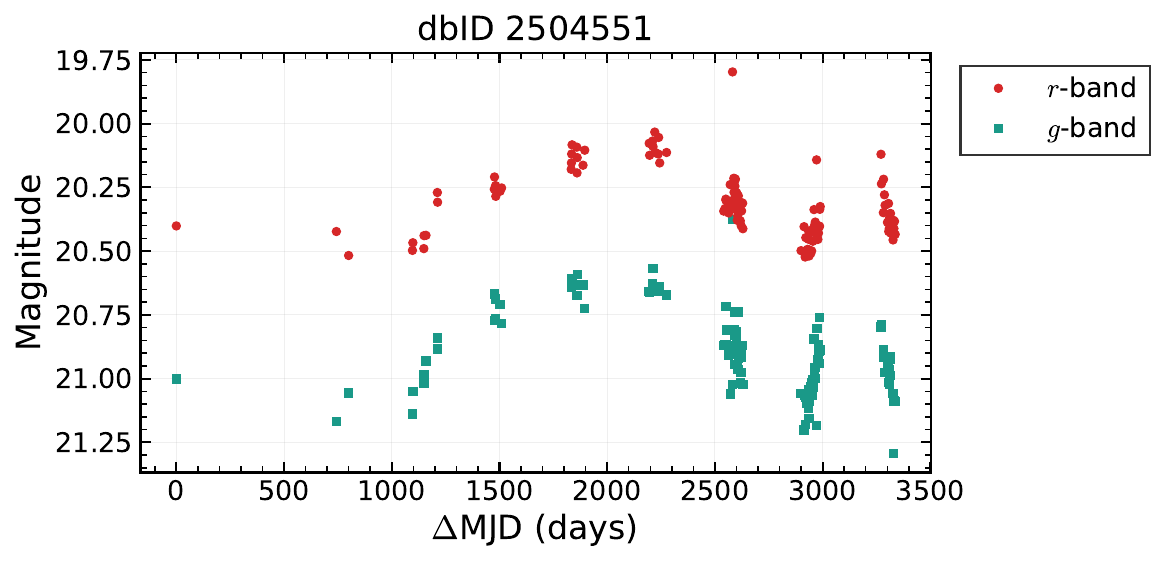}
      \hfill
      \includegraphics[width=0.48\columnwidth]{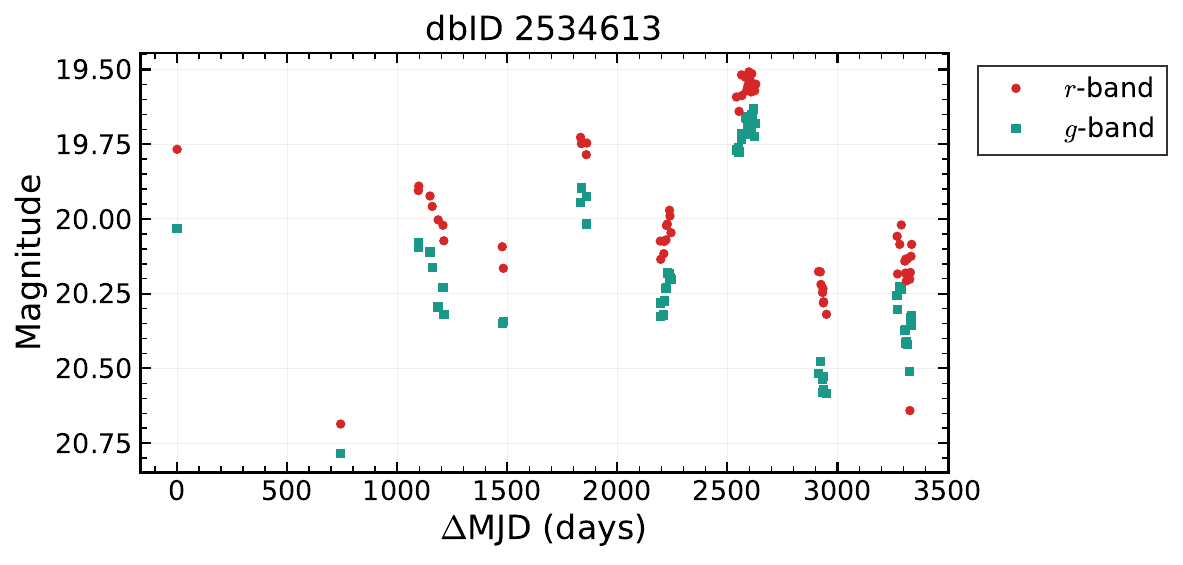}
      \caption{continued.}
  \end{figure}

  \begin{figure}[H]\ContinuedFloat
      \centering
      \includegraphics[width=0.48\columnwidth]{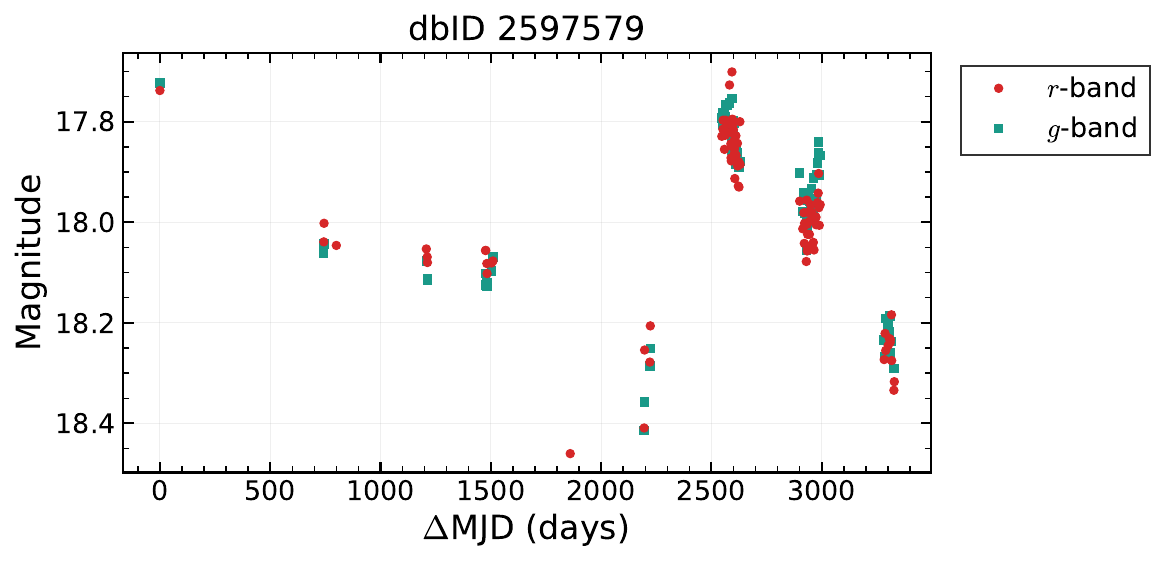}
      \hfill
      \includegraphics[width=0.48\columnwidth]{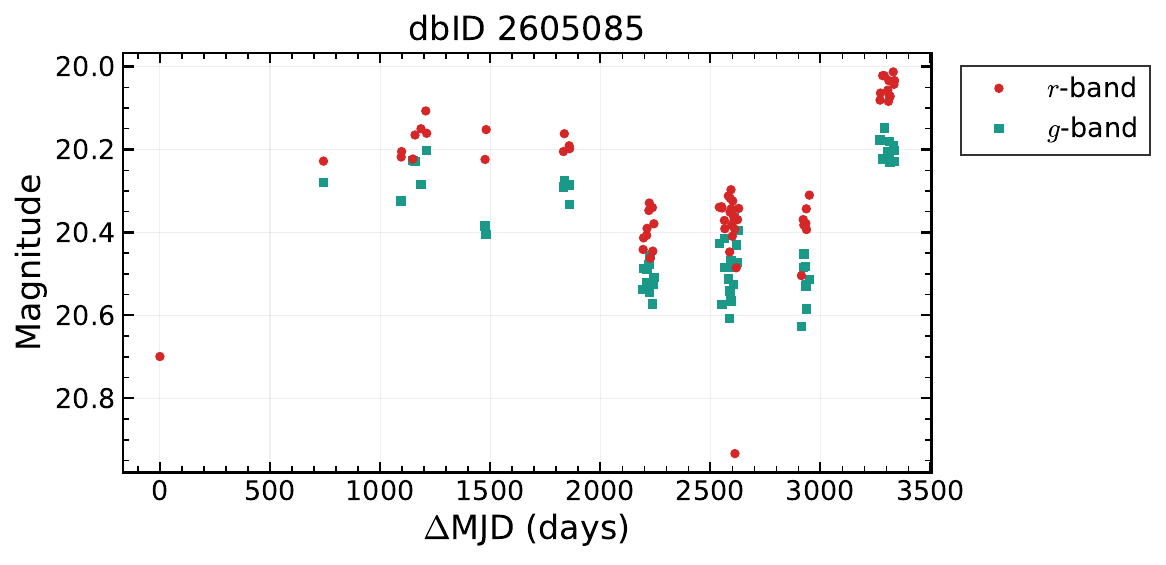}
      \caption{continued.}
  \end{figure}

  \begin{figure}[H]\ContinuedFloat
      \centering
      \includegraphics[width=0.48\columnwidth]{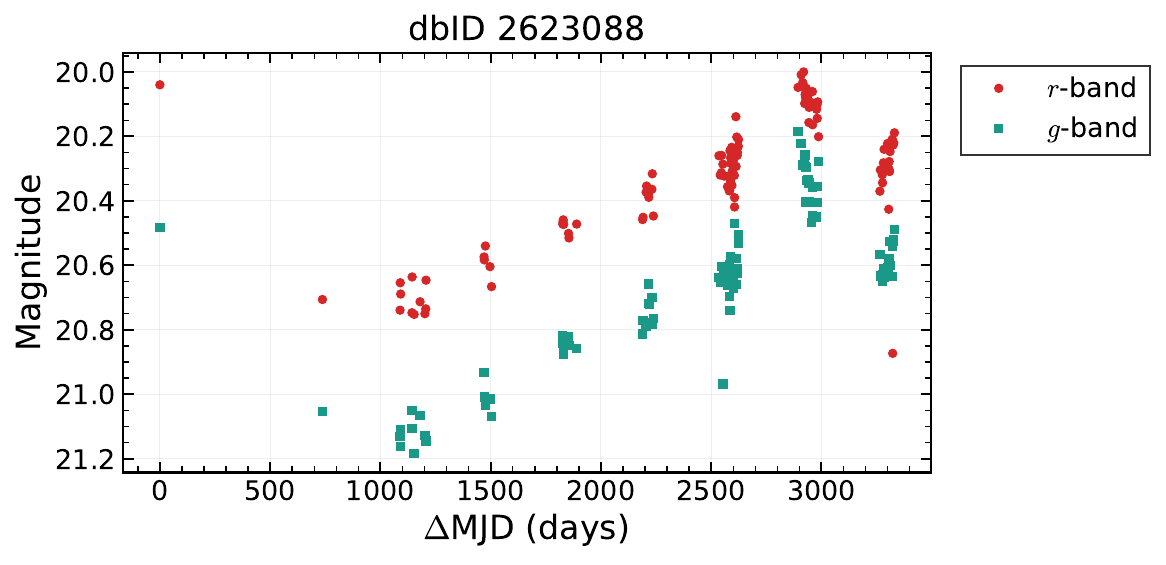}
      \hfill
      \includegraphics[width=0.48\columnwidth]{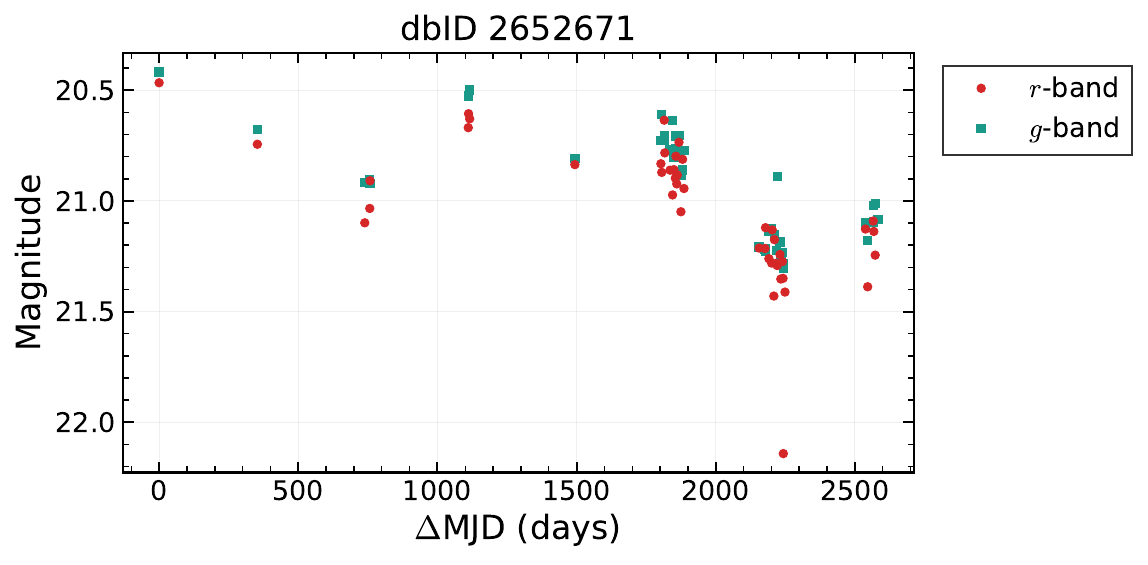}
      \caption{continued.}
  \end{figure}

  \begin{figure}[H]\ContinuedFloat
      \centering
      \includegraphics[width=0.48\columnwidth]{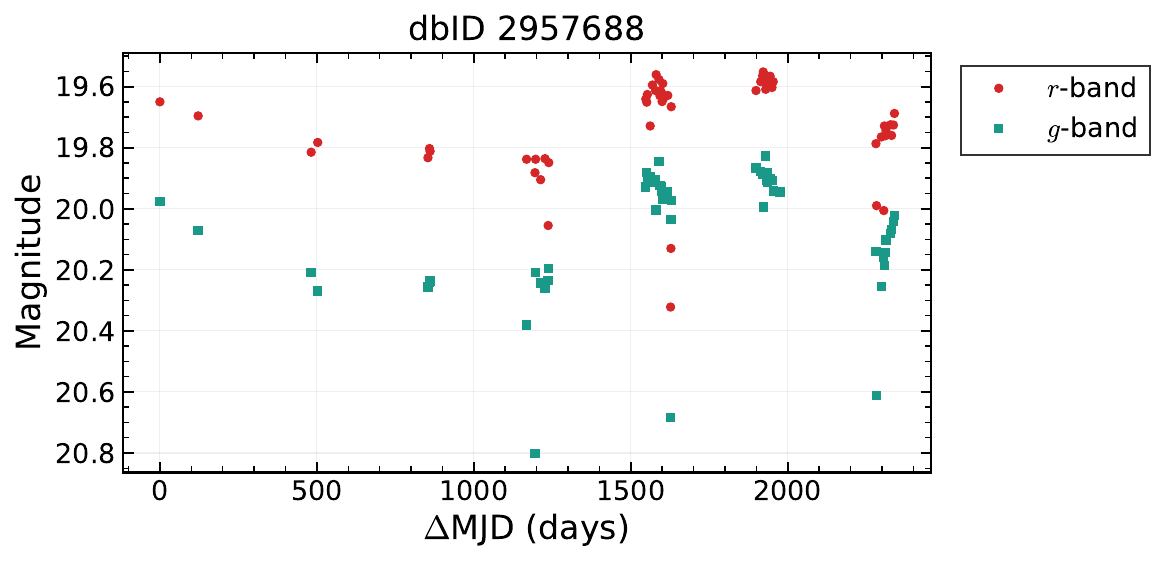}
      \hfill
      \includegraphics[width=0.48\columnwidth]{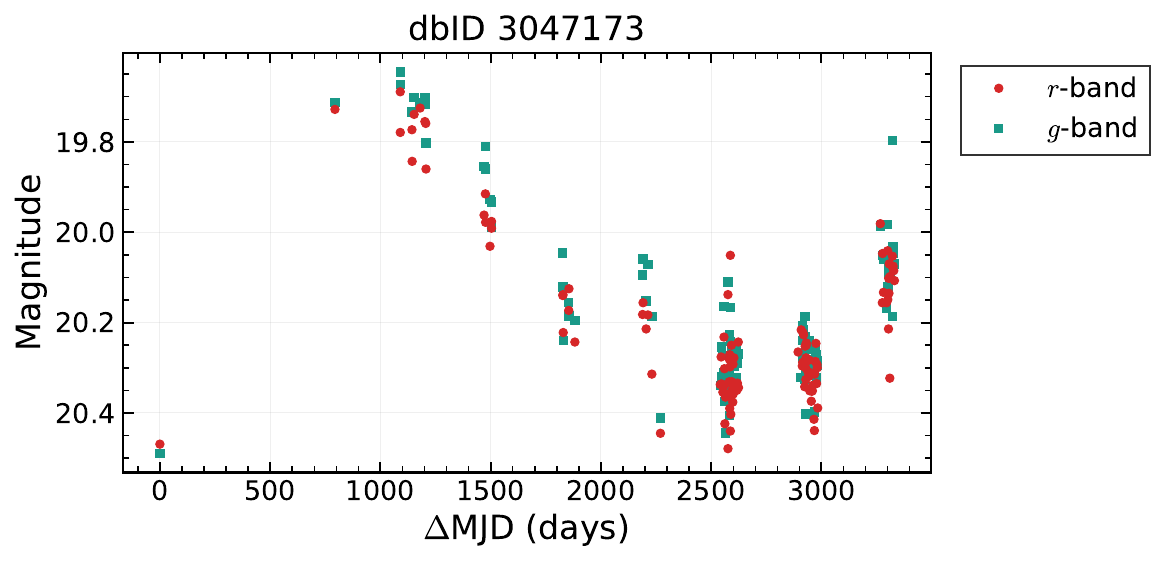}
      \caption{continued.}
  \end{figure}

  \begin{figure}[H]\ContinuedFloat
      \centering
      \includegraphics[width=0.48\columnwidth]{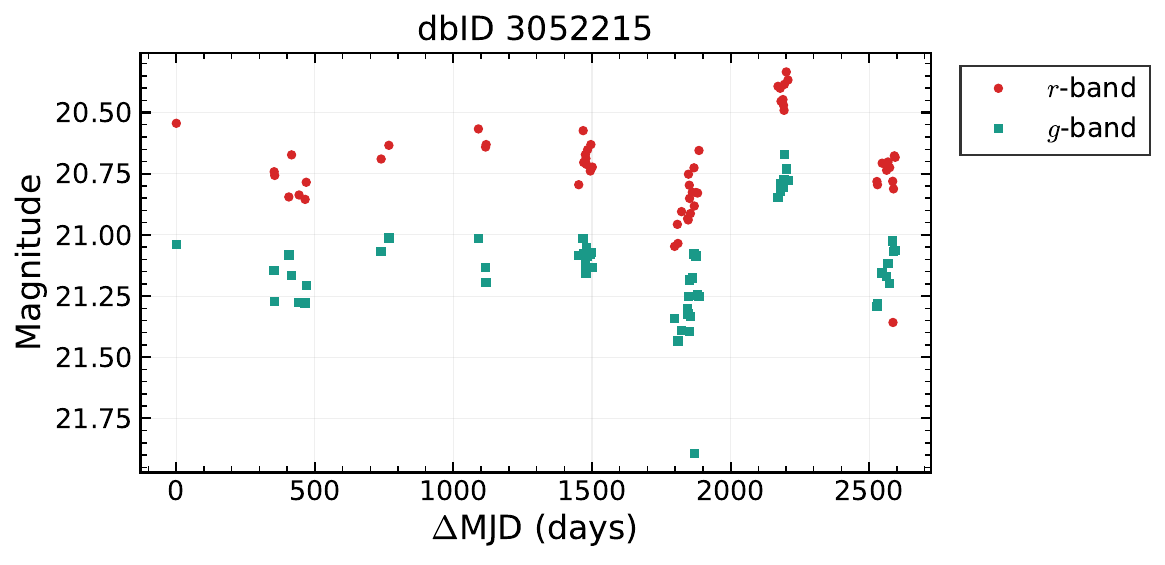}
      \hfill
      \includegraphics[width=0.48\columnwidth]{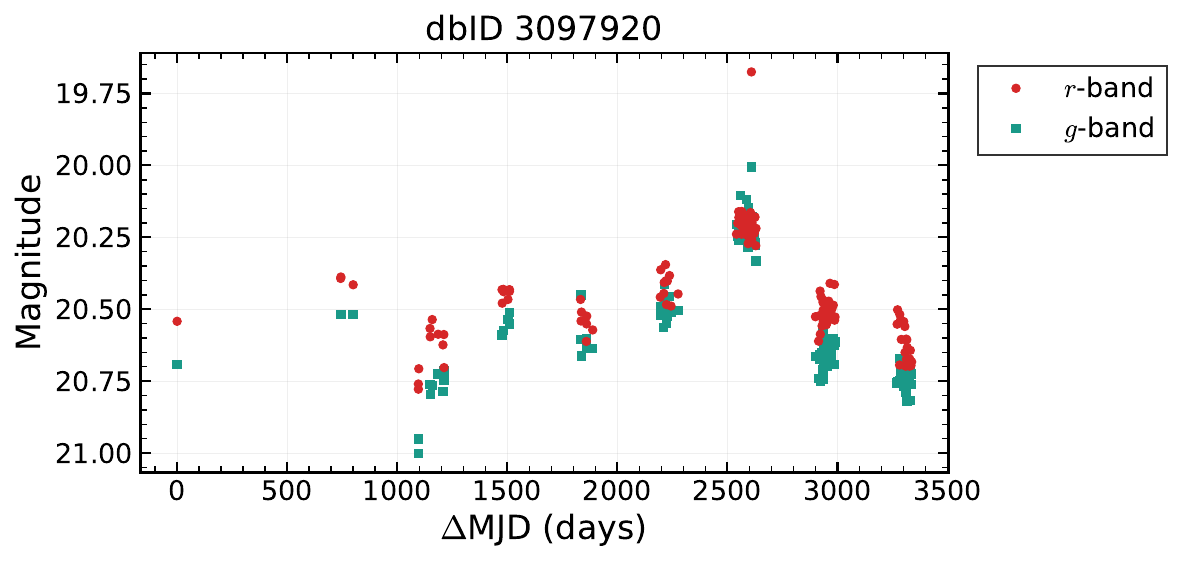}
      \caption{continued.}
  \end{figure}

  \begin{figure}[H]\ContinuedFloat
      \centering
      \includegraphics[width=0.48\columnwidth]{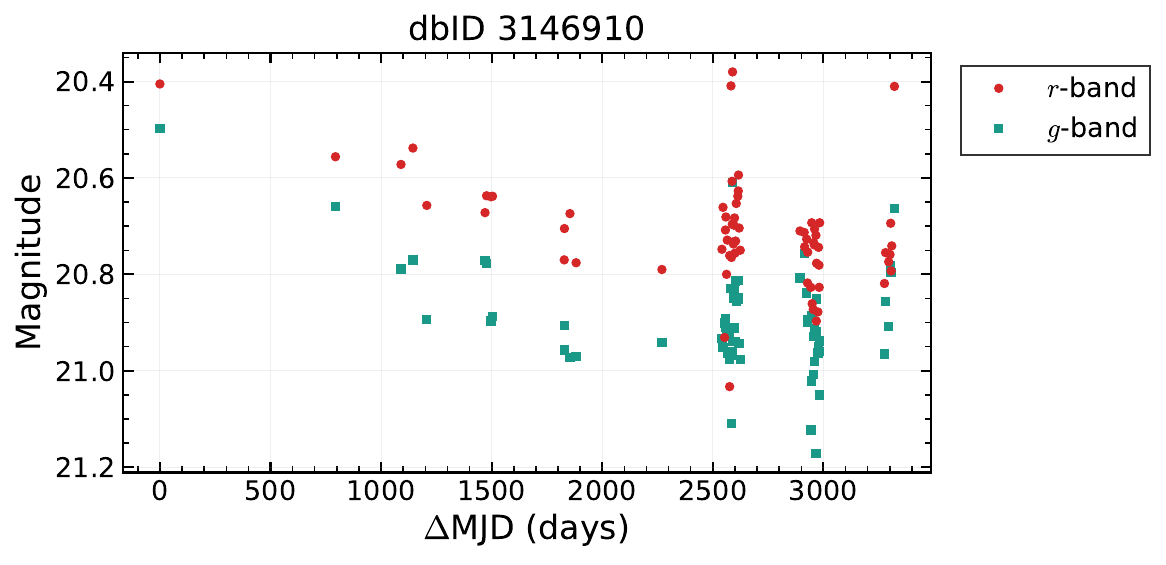}
      \hfill
      \includegraphics[width=0.48\columnwidth]{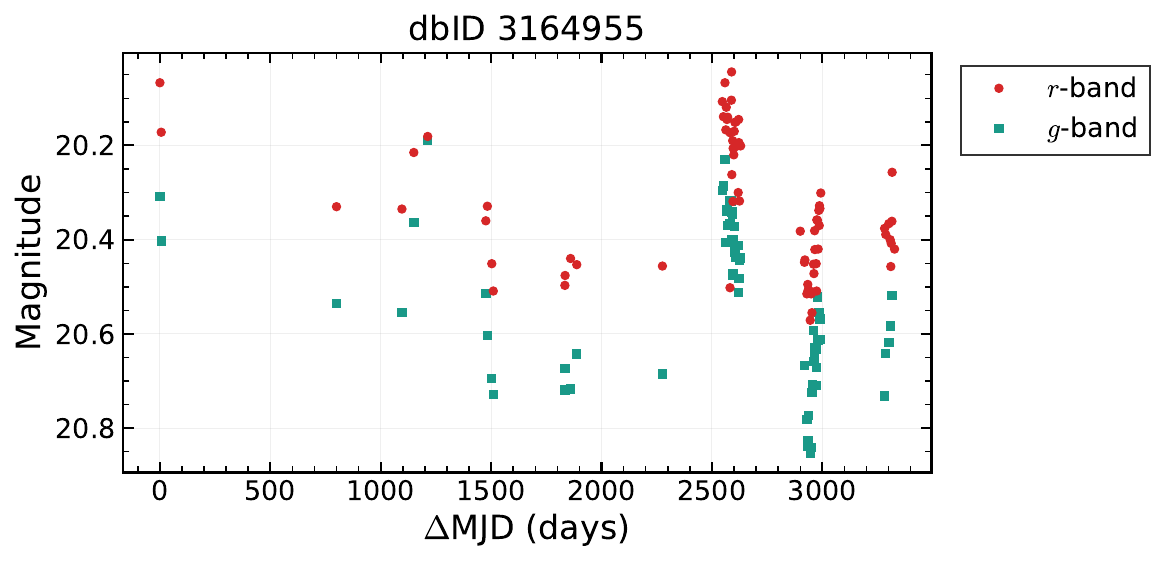}
      \caption{continued.}
  \end{figure}

  \begin{figure}[H]\ContinuedFloat
      \centering
      \includegraphics[width=0.48\columnwidth]{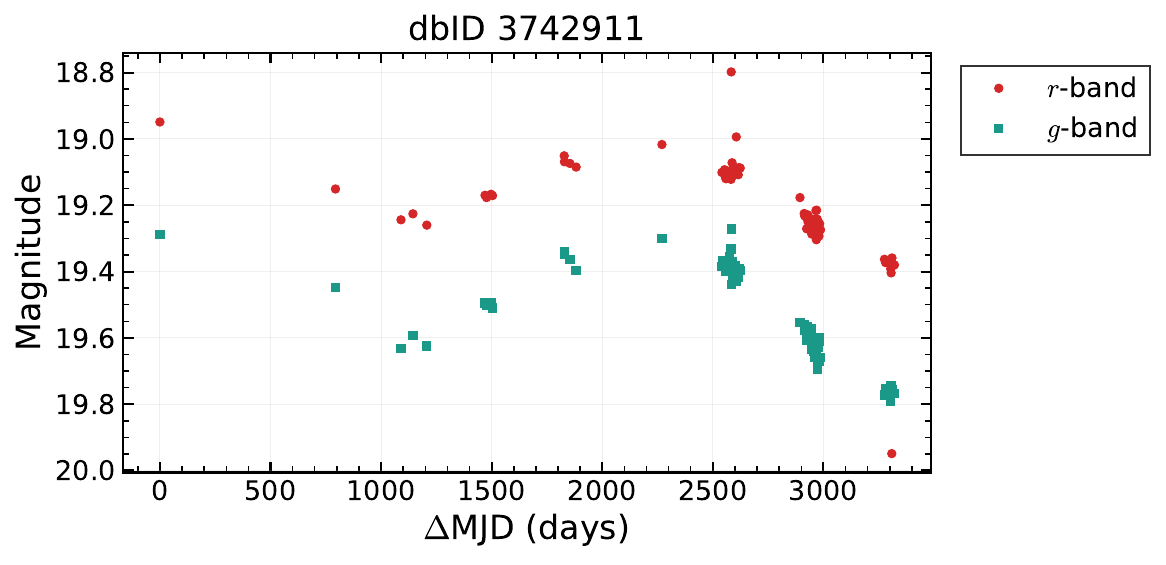}
      \hfill
      \includegraphics[width=0.48\columnwidth]{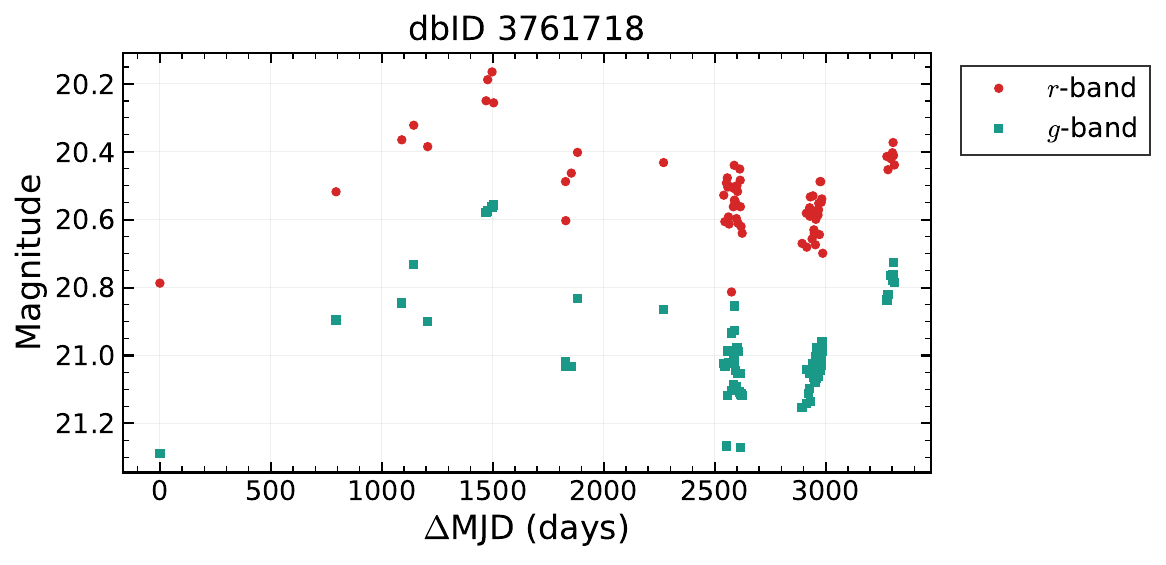}
      \caption{continued.}
  \end{figure}

  \begin{figure}[H]\ContinuedFloat
      \centering
      \includegraphics[width=0.48\columnwidth]{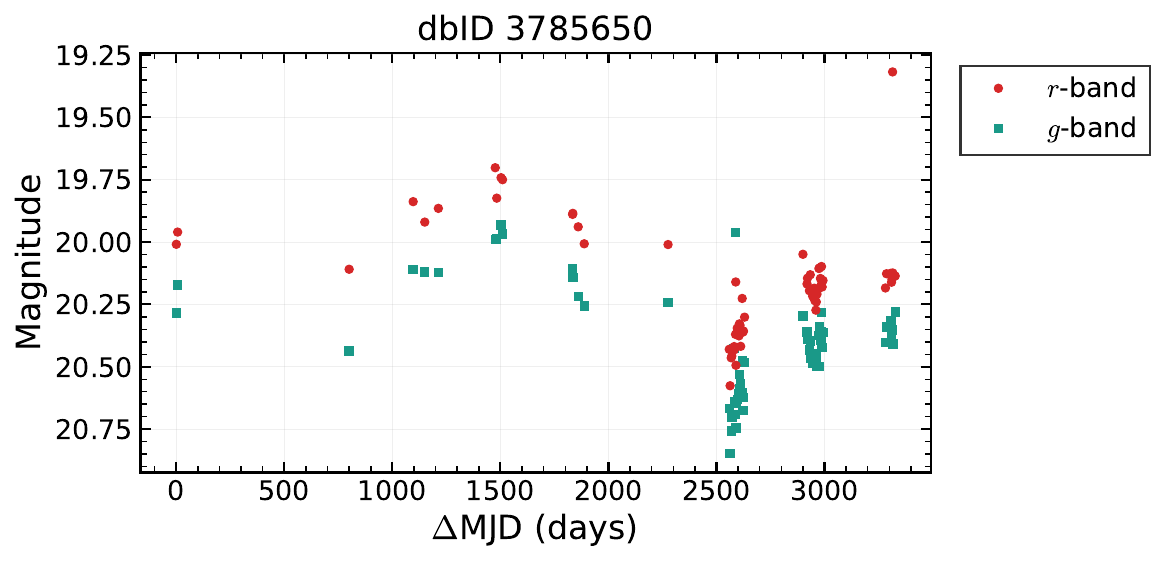}
      \hfill
      \includegraphics[width=0.48\columnwidth]{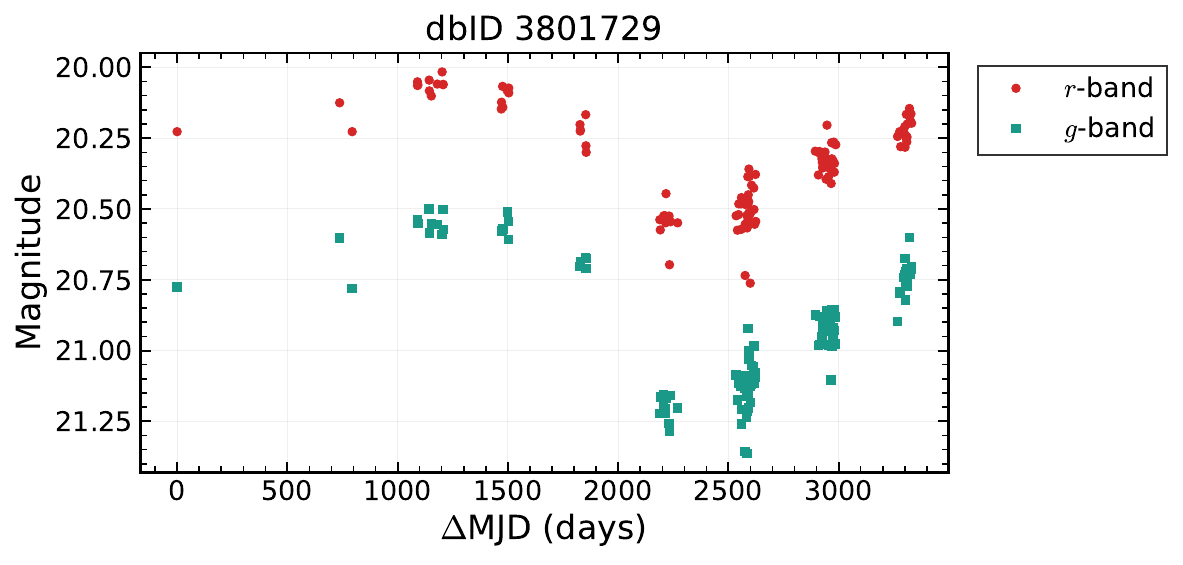}
      \caption{continued.}
  \end{figure}

  \begin{figure}[H]\ContinuedFloat
      \centering
      \includegraphics[width=0.48\columnwidth]{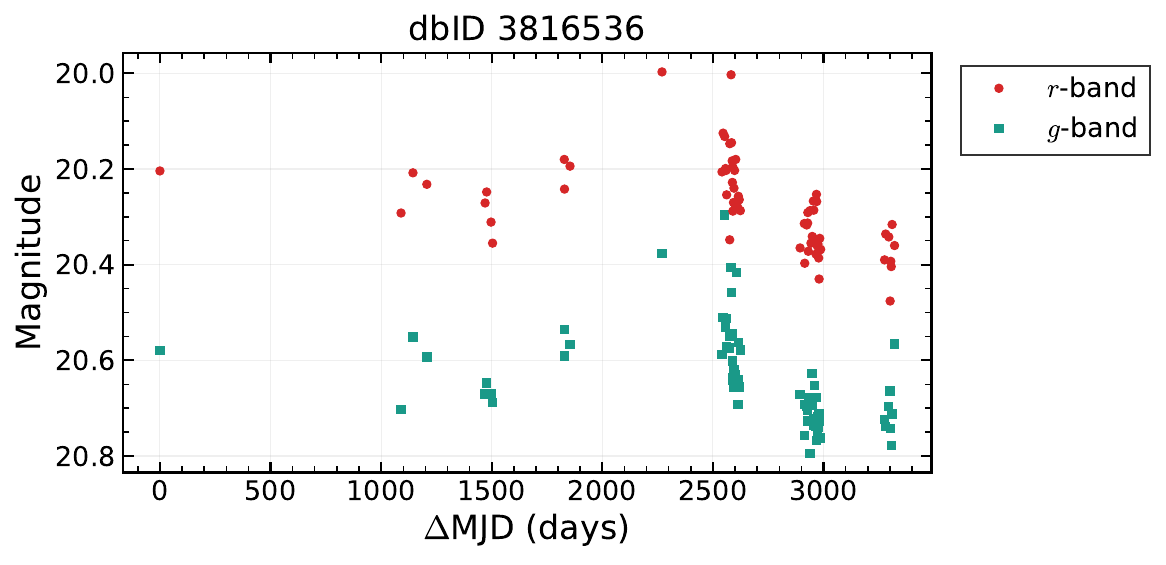}
      \hfill
      \includegraphics[width=0.48\columnwidth]{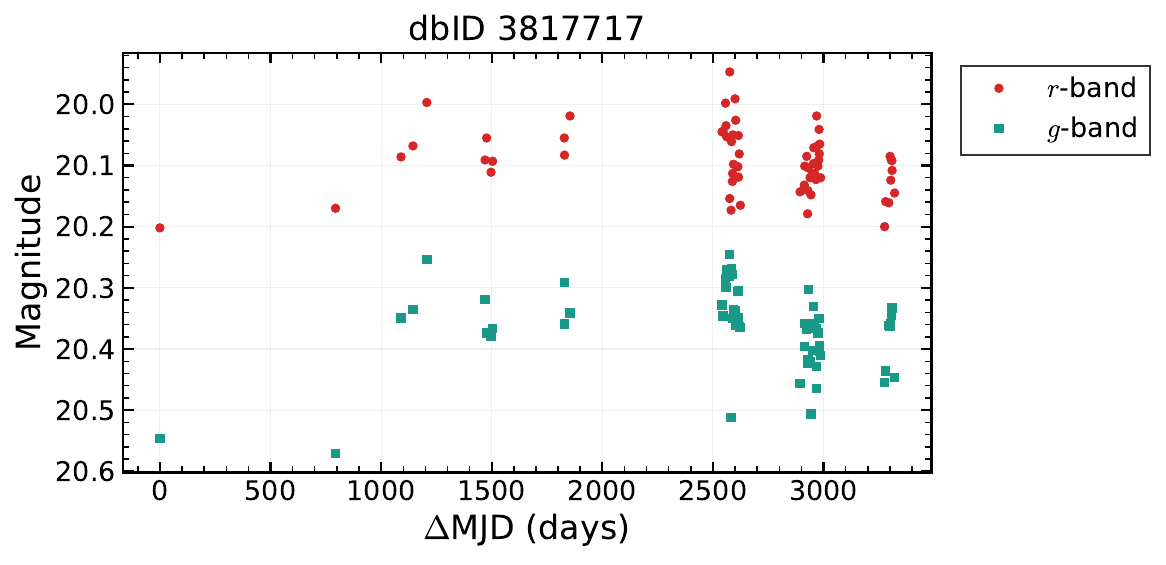}
      \caption{continued.}
  \end{figure}

  \begin{figure}[H]\ContinuedFloat
      \centering
      \includegraphics[width=0.48\columnwidth]{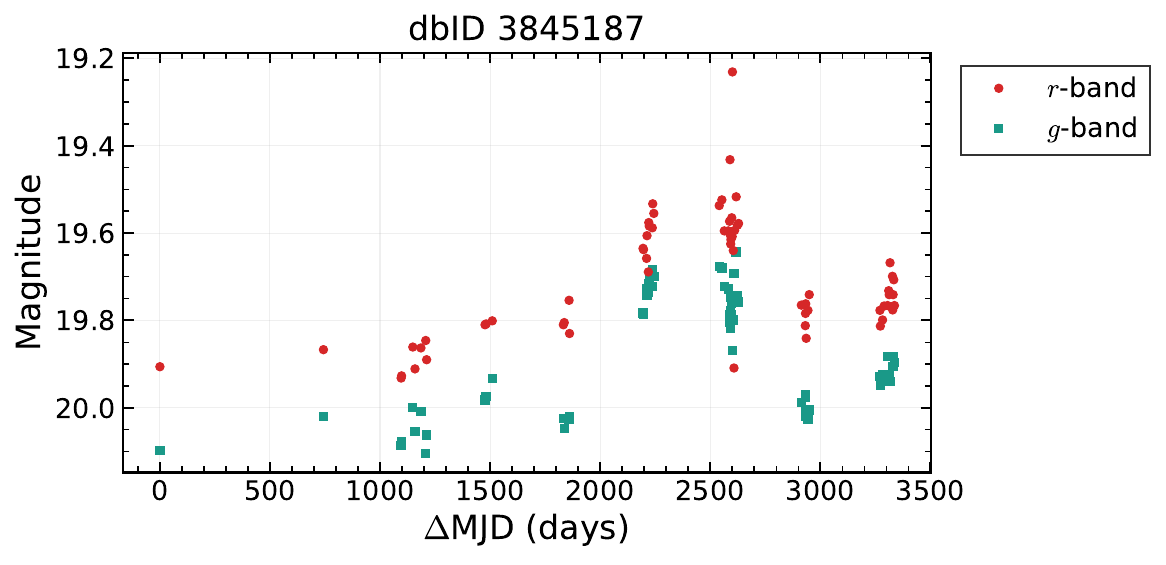}
      \hfill
      \includegraphics[width=0.48\columnwidth]{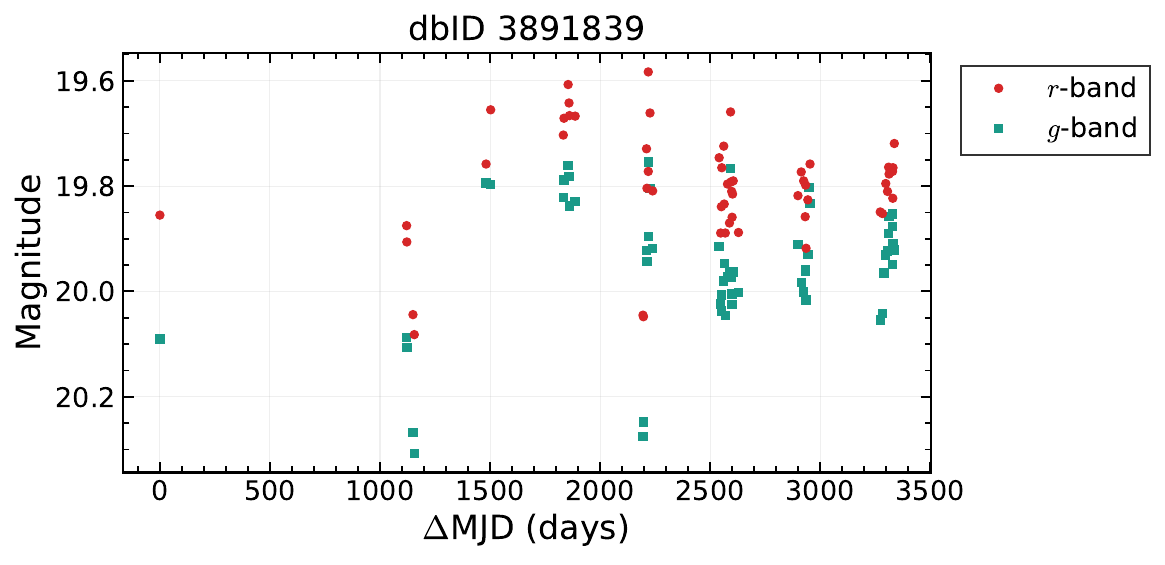}
      \caption{continued.}
  \end{figure}

  \begin{figure}[H]\ContinuedFloat
      \centering
      \includegraphics[width=0.48\columnwidth]{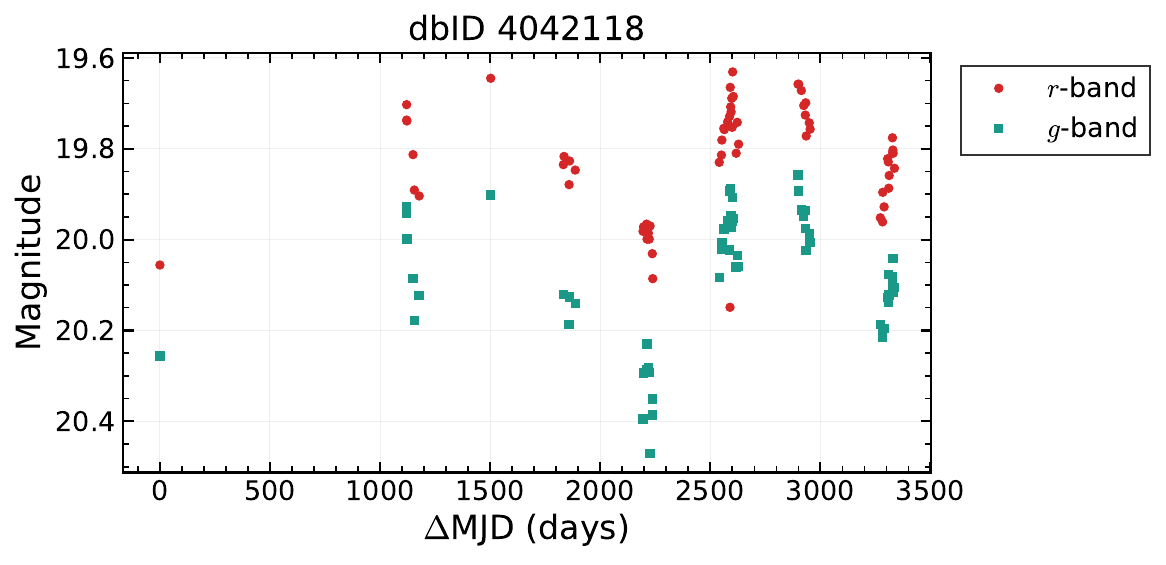}
      \hfill
      \includegraphics[width=0.48\columnwidth]{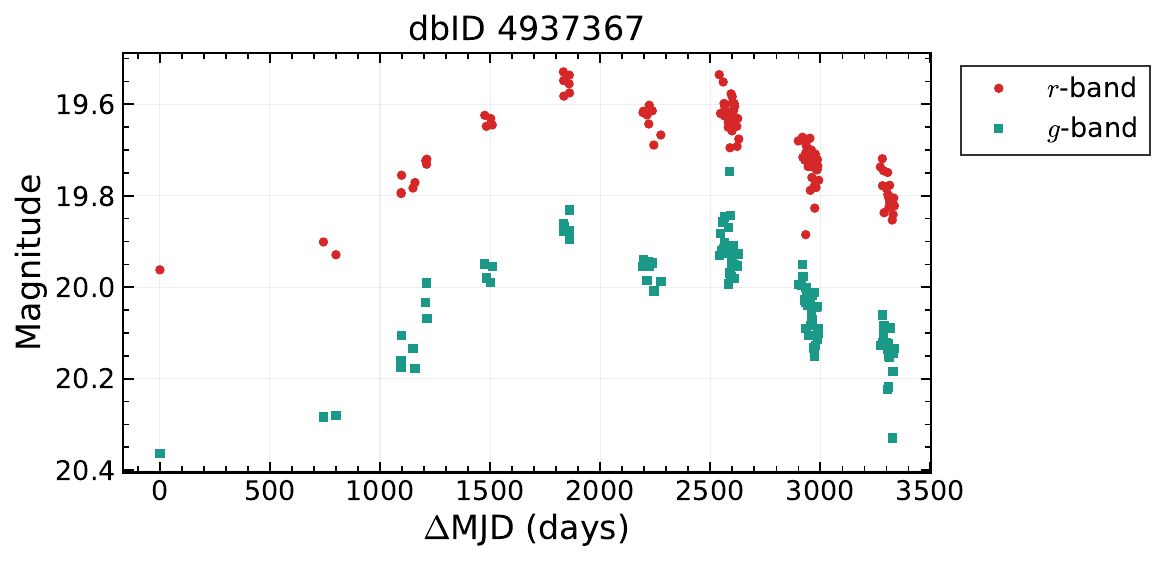}
      \caption{continued.}
  \end{figure}

  \begin{figure}[H]\ContinuedFloat
      \centering
      \includegraphics[width=0.48\columnwidth]{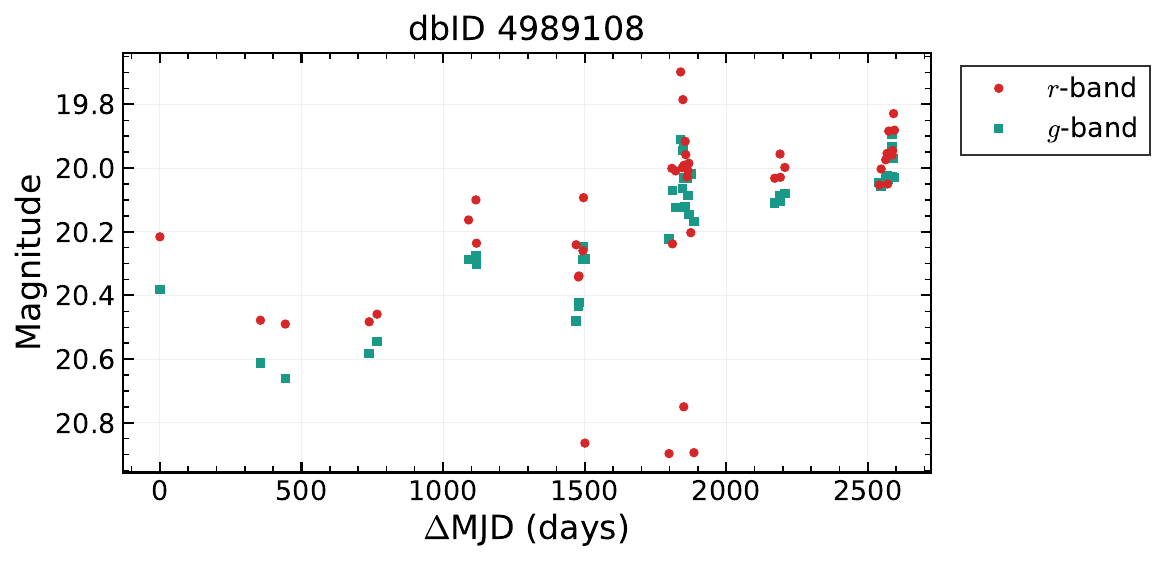}
      \caption{continued.}
  \end{figure}

\end{appendix}

\end{document}